\documentclass[aps,prf,preprint,groupedaddress]{revtex4-2}

\usepackage{graphbox}
\usepackage{xcolor}
\newcommand{\berengere}[1] {{\color{black} {#1}}}

\newcommand{\xx}{\underline{x}}
\newcommand{\uu}{\underline{u}}
\newcommand{\pphi}{\underline{\phi}}
\newcommand{\hh}{\hspace{3.5cm}}

\begin{document}



\title{Proper Orthogonal Decomposition Analysis and Modelling
of the wake deviation behind a squareback Ahmed body}


\author{B\'ereng\`ere Podvin}
\affiliation{LIMSI, CNRS, Universit\'e Paris-Saclay}
\author{St\'ephanie Pellerin}
\affiliation{LIMSI, CNRS, Universit\'e Paris-Saclay}
\author{Yann Fraigneau}
\affiliation{LIMSI, CNRS, Universit\'e Paris-Saclay}
\author{Antoine Evrard}
\affiliation{ENSTA}
\author{Olivier Cadot}
\affiliation{University of Liverpool}


\date{\today}

\begin{abstract}
We investigate numerically the 3-D flow around
a squareback Ahmed body at  Reynolds number $Re=10^4$. 
Proper Orthogonal Decomposition (POD) is applied to 
\berengere{ a symmetry-augmented database} in order 
to describe and model the flow dynamics. 
Comparison with experiments at a higher Reynolds number 
in a  plane section  of the near-wake at mid-height shows that the simulation 
captures several features of the experimental  flow, in particular
the antisymmetric quasi-steady deviation mode.
3-D POD analysis allows us to classify the different physical processes in terms
of mode contribution to the  kinetic energy over the entire domain. It is found that the dominant fluctuating mode
on the entire domain corresponds to the 3-D quasi-steady wake deviation, and  that its amplitude
 is well estimated from  2-D near-wake data.  
The next most energetic flow fluctuations
consist  of vortex shedding and bubble pumping mechanisms. It is found 
that the amplitude of the deviation is negatively correlated with
the intensity of the vortex shedding in the spanwise direction and the suction drag coefficient.
Finally, we find that despite the slow convergence of the decomposition, a 
POD-based low-dimensional model reproduces the dynamics of the wake
deviation observed experimentally, as well as the main 
characteristics of the  global modes identified in the simulation.  
\end{abstract}

\maketitle

\section{Introduction}

A surprisingly generic feature of the flow around 
\berengere{symmetric} bodies
at high Reynolds numbers is the presence of permanent 
symmetry-breaking structures in the wake. 
These have been observed for the sphere 
\cite{kn:grandemange14}, bullet \cite{kn:rigas14} 
, the flat plate
\cite{kn:cadot16}, as well as academic models for ground  vehicles such as
the Ahmed body \cite{kn:grandemange13} and the Windsor body \cite{kn:perry16}. 
The origin and dynamics of these structures 
has not been entirely elucidated,
although they have been the object of several experimental 
and numerical studies.
They appear to be connected to the first bifurcation observed
at much lower Reynolds number (see \cite{kn:fabre08},  
\cite{kn:grandemange12}). 
At higher Reynolds numbers, rapid switches between different quasi-stable states can be 
observed, as described by \cite{kn:rigas15_a} and \cite{kn:brackston16}.

In the remainder of the paper, we will focus on the Ahmed body. 
 Grandemange {\it et al.} \cite{kn:grandemange13} have established that
the appearance of bi- or multi-stable states for the flow around a squareback Ahmed body 
depends on the value of the ground clearance and the aspect ratio of the body base. 
Pasquetti and Peres \cite{kn:pasquetti15} carried out the first numerical simulation of the Ahmed body
that was able to reproduce the steady wake deviation.
However switches were not observed in Pasquetti and Peres's simulation, since the typical time
separating two switches is on the order of 1000 $U/H$ where $U$ is the incoming flow speed and $H$ the body characteristic height
(see \cite{kn:lucas17}).
In more recent work Dalla Longa {\it et al.} \cite{kn:dallalonga19} 
were able to integrate over sufficiently long times to capture the switch in the wake deviation.
However they could only observe  one switch, due to the still longer time scale
separating two switches. 
They also applied Proper Orthogonal Decomposition and Dynamic Mode Decomposition (DMD, \cite{kn:schmid10}) to extract
the dynamics of the flow.
Interaction of the near-wake recirculation zone 
with the surrounding shear layers is expected to play a part in the occurence of switches.
They proposed that the switch is triggered by large hairpin vortices. 

Volpe {\it et al.} \cite{kn:volpe15} carried out a spectral analysis
of the wake behind the squareback Ahmed body.
 Three low frequencies were identified in their experimental data: 
vortex-shedding modes with  Strouhal numbers of 0.13 and
0.19 in the far wake, and wake pumping motion at a 
Strouhal number of 0.08 in the recirculation zone. 
Pavia {\it et al.} \cite{kn:pavia18} have applied  POD
to both the experimental pressure and velocity field  of a Windsor body to identify structures corresponding
to the bi-stable modes and the dynamics of the switch.
They identified pumping motion in the wake
with a bi-stable vortical structure in the streamwise direction
and derived a phase-averaged model that describes the  switch between
the quasi-stable states. 
They confirmed the observation made
by Evrard {\it et al.} \cite{kn:evrard16}
and Cadot {\it et al.} \cite{kn:cadot16}  
that the symmetric state corresponds to a lower drag.

Drag reduction through control of the flow asymmetry has been the object of several passive
and active control strategies
\cite{kn:lucas17}, \cite{kn:wassen10}, \cite{kn:brackston16}, \cite{kn:li16},
\cite{kn:varon17}, \cite{kn:evstafyeva17}, \cite{kn:rigas15_b}.
 An important question is to determine 
how the different structures present in the flow contribute
to the drag.
A low-dimensional description and modelling  of the large-scale structures 
could be  beneficial for understanding, predicting and 
ultimately controlling the flow dynamics. 
Reduced-order models were developed by  
\cite{kn:brackston16}, \cite{kn:rigas15_a} 
in order to control the deviation of the wake. 

In the present work,  we  aim to provide a large-scale description of
the flow by  applying 
Proper Orthogonal Decomposition
to a numerical investigation of the flow around a squareback Ahmed body 
at a Reynolds number of $Re=10^4$.
The spanwise to vertical aspect ratio of the body is 1.18, so that
bistability corresponds to two  asymmetric states which are symmetric through reflection of the vertical mid-plane.
Due to the long time scales separating switches, the change in the wake deviation could not be captured by the simulation.  However, reflection symmetry was enforced artificially, so that we provide a description of the structures of the flow as a superposition of reflection-symmetric and
reflection-antisymmetric modes.

The paper is organized as follows: section 2 
presents the 3-D numerical configuration, while section 3 describes
the specifics of  Proper Orthogonal Decomposition.  
Section 4 presents a 2-D comparison of the numerical simulation and
an experimental configuration corresponding to the same geometry
but a higher Reynolds number,
section 5 presents a 3-D POD analysis of the structures in the 
full configuration; 
a POD-based low-dimensional model is constructed in section 6 and compared with experimental and numerical results. 
A conclusion is given in section 7. 

\section{Numerical configuration}

\begin{figure}
\hspace{-2.5in}
\includegraphics[height=70mm]{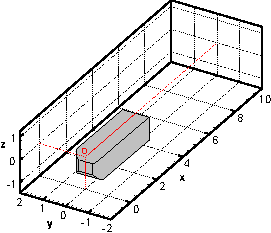}  
\caption{Numerical configuration. The Ahmed body and the recirculation zone behind it
are idenfied by a surface of zero streamwise velocity. 
The streamwise, spanwise and wall-normal positions are referred to as
$x$, $y$, $z$. \berengere{The length, width and height of the body 
are respectively $L$, $W$ and $H$.} 
The origin of the axes is taken
at the top and foremost position  of the Ahmed body in the vertical
symmetry plane. } 
\label{fig1:config} \end{figure}

Figure ~\ref{fig1:config} presents the numerical configuration.
The dimensions of the squareback Ahmed  body are the same as the
experimental configuration of Evrard {\it et al.} \cite{kn:evrard16} i.e.
$L=1.124m, H=0.297m, W=0.35m$. 
The  ground clearance
\berengere{ 
(distance from the body to the lower boundary of the 
domain)  }
 is \berengere{0.3H} in the simulation.
The Reynolds number based on the incoming velocity $U$, body height $H$
and fluid viscosity is $10^4$. 
The foremost and upper part body defines the reference position $(x=0, z=0)$.
The plane $y=0$ corresponds to the mid-plane of the body. 
The domain extent   in the streamwise (x), spanwise (y) and vertical (z)  directions is
$  [-1, 10]H \times [-2, 2]H \times [-1.3, 1.2]H$. 

The code used is SUNFLUIDH, an in-house code developed at LIMSI based on 
a second-order finite volume approach which has been described for instance 
in \cite{kn:podvinpf17}. 
The temporal discretization is based on a second-order  backward Euler scheme. Diffusion terms 
are treated implicitly and convective terms are solved with an Adams-Bashforth scheme. The 
Poisson equation for the computation of pressure field 
is solved iteratively.
We use $(512 \times 256 \times 256)$ grid points in respectively
the longitudinal direction $x$, 
the spanwise direction $y$,
and the vertical direction $z$. The Cartesian grid is refined 
close to surfaces. Periodic boundary conditions
are used in the spanwise direction. No-slip velocity 
boundary conditions on the Ahmed body \berengere{and on the  ground}
are implemented by adjusting the size of the loops.  
The simulation was initialized from a uniform condition. 
About 100 time units based on the upstream velocity $U$ 
and body height $H$ were necessary for the flow to develop and statistical convergence to be reached.
We note that all times will be expressed in those units in the remainder of the paper. All lengths will be made nondimensional with the body height $H$. 

\section{Proper Orthogonal Decomposition}

The main tool of analysis used in this paper is Proper 
Orthogonal Decomposition (POD) \cite{kn:lumleyPOD}.
The field  on a domain $\Omega$ is written as a superposition of spatial modes
\begin{equation} 
\uu(\xx,t) = \sum_{n} \tilde{a}_n(t) \pphi_n(\xx) 
\end{equation}
where the modes are orthogonal (and can be made orthonormal), i.e
\[ \int \pphi_n(\xx) . \pphi_m(\xx) d\xx = \delta_{nm}. \]
The modes  can be ordered by decreasing energy $\lambda_1 \ge \lambda_2 \ge \ldots \ge \lambda_n = <\tilde{a}_n \tilde{a}_n >$, where
$<.>$ represents a time average.    
The amplitudes $\tilde{a}_n$ can be obtained by projection of the velocity field
onto the spatial modes:
\begin{equation} 
\tilde{a}_n(t)= \int_{\Omega} \uu(\xx,t). \pphi_n(\xx) d\xx.  
\end{equation} 
In the remainder of the paper we will consider normalized amplitudes
$a_n= \tilde{a}_n/\sqrt{\lambda_n}.$

In all that follows POD is implemented following the method of snapshots \cite{kn:siro87}
which is based on computing the autocorrelation between the different samples
of the field obtained at times $t_i, i=1, \ldots, N$:
\[ C_{ij}= \frac{1}{N} \int_{\Omega} \uu(\xx,t_i).\uu(\xx',t_j) d\xx, \]
and extracting the eigenvalues $\lambda_n$ and temporal eigenvectors $A_{in}=a_n(t_i)$
such that
\begin{equation}
C A = \lambda A.
\end{equation}
The spatial modes can then be reconstructed using
\begin{equation}
\tilde{\pphi}_n(\xx) =  u(\xx, t_i) A_{in} 
\end{equation}
and renormalizing ${\pphi}_n(\xx) = \frac{1}{N_n} \tilde{\pphi}_n(\xx) $  
where
\begin{equation}
N_n^2 = \int \tilde{\pphi}_n(\xx) .  \tilde{\pphi}_n(\xx) d \xx.  
\end{equation}

Different spatial domains, as well as different quantities 
will be considered in the decomposition.
In section 4, we will limit the analysis to 2-D velocity fields
in the near-wake region in order to match the data 
of Evrard {\it et al.} \cite{kn:evrard16}.
In section 5, we will apply POD to the full 3-D velocity field to the full numerical domain. 

\begin{figure} 
\hspace{-2in}
\begin{tabular}{ll}
\includegraphics[height=30mm]{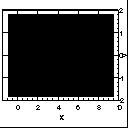} & 
\includegraphics[height=30mm]{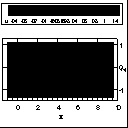}  
\\
\end{tabular}
\caption{  Streamwise velocity contours of the mean flow; 
left) horizontal plane at mid-height $z=-0.5$;
right) vertical mid-plane at $z=0$.}
 \label{fig2:mean}
\end{figure}

A key ingredient of the procedure is the definition of the data set.
Cross-sections of the time-averaged streamwise velocity 
in the mid-height plane $z=-0.5$ and the vertical mid-span plane $y=0$ 
are represented in figure ~\ref{fig2:mean}. 
We observe a steady deviation of the wake
in agreement with previous observations \cite{kn:pasquetti15}.
This deviation is breaking the reflection symmetry with respect
to the vertical mid-plane.
However the equations are symmetric 
- for each flow realization, the flow obtained by symmetry with respect
to the vertical midplane is also a possible solution.
Following recommendations in ~\cite{kn:HLB}, 
an enlarged dataset   enforcing the statistical symmetry could then be created 
as follows: for each snapshot of the original dataset,  a symmetrized 
snapshot corresponding to the image of the snapshot by the reflection symmetry 
was created.  The size of the new data set was therefore twice that of the original one.
By construction, POD modes are thus either symmetric or antisymmetric
with respect to the vertical midplane $y=0$, 
\berengere{so that the amplitude of a POD mode
for a given snapshot of the original dataset is either identical or opposite to that on the symmetrized snapshot}.
This allows us to break down  flow patterns into symmetric and antisymmetric components.

We emphasize that individual POD modes are different from coherent structures.
A coherent structure typically corresponds to  local patterns identified in a realization.
POD modes are defined over the full domain chosen for decomposition, and 
 every single realization contains a combination of modes.
A coherent structure will therefore correspond to a combination of a few POD modes,
which may be restricted to a portion of a spatial domain.
As a consequence of  the data enlargement procedure we have adopted,
the mean flow of the simulation does not correspond to one single mode
but to the sum of the first two modes representing respectively
the symmetric and the antisymmetric part of the mean flow.

\section{2-D POD: Comparison with the experiment}

In this section we apply POD analysis in the near-wake
to the same variables defined over the same domain in both the experiment
and the numerical simulation, i.e. the streamwise and the spanwise velocity components $u$ et
$v$ defined over the domain $ L \le x \le L + 1.2H$, $ -0.6 \le y \le 0.6H$.
Details are indicated in table ~\ref{tab:comparison}.
The main differences are \\
(i) the Reynolds numbers considered, as $Re=4$ $10^5$ in Evrard's experiment and $Re=10^4$ in our numerical simulation \\ 
(ii) the time resolution - which is 
$25 H/U$ in the experiment versus 0.5H/U in the simulation.

\begin{table}[h]
\begin{tabular}{cccc}
Type & Re &  $\Delta T$ & snapshot number (original data set)  \\  
Simulation & $10^3$ &  0.5 & 150  \\ 
Experiment & $4$ $10^5$ & 25.25 & 400  \\ 
\end{tabular}
\caption{ Comparison between model and experiment}
\label{tab:comparison}
\end{table}

\begin{figure}[h]
\hspace{-3in}
\includegraphics[height=70mm]{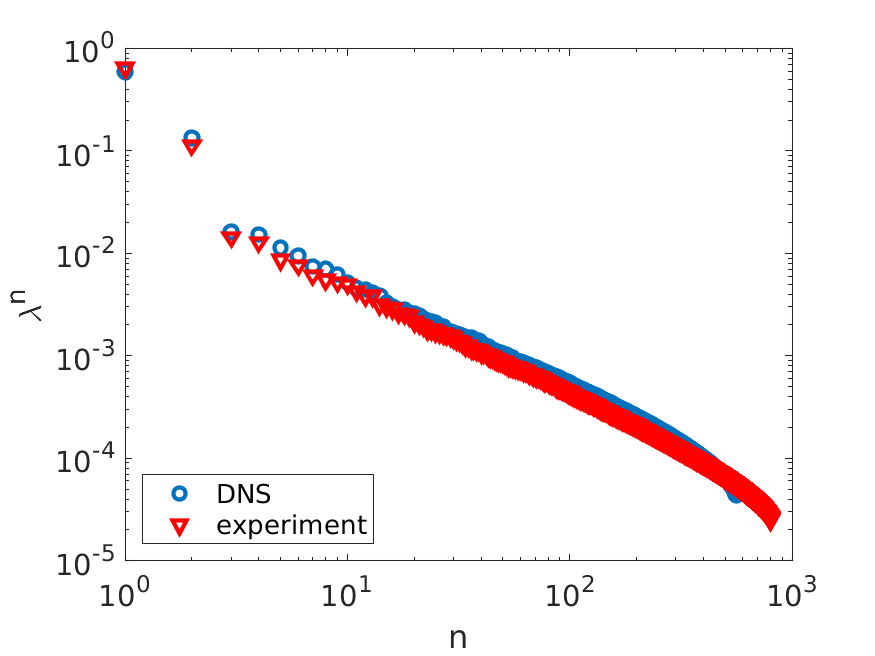}  
 \caption{  Near-wake 2-D POD spectrum in the simulation and experiment.}
 \label{fig3:lambda2d}
\end{figure}

\begin{figure}[h]
\hspace{-2in}
\begin{tabular}{cccc}
\hh n=1 &\hh n=2 &\hh n=3 &\hh n=4 \\
\includegraphics[height=30mm]{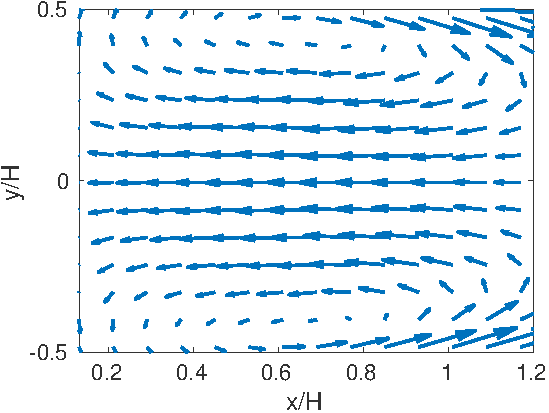} & 
\includegraphics[height=30mm]{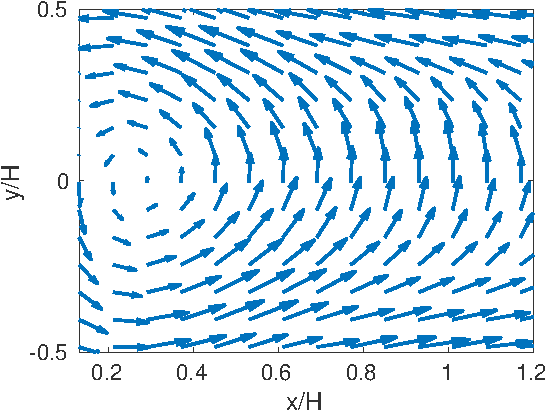} & 
\includegraphics[height=30mm]{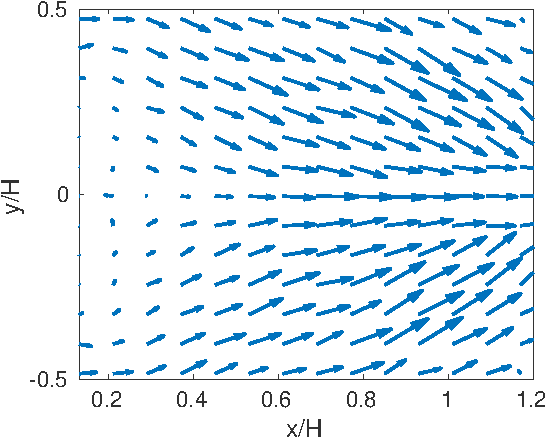} & 
\includegraphics[height=30mm]{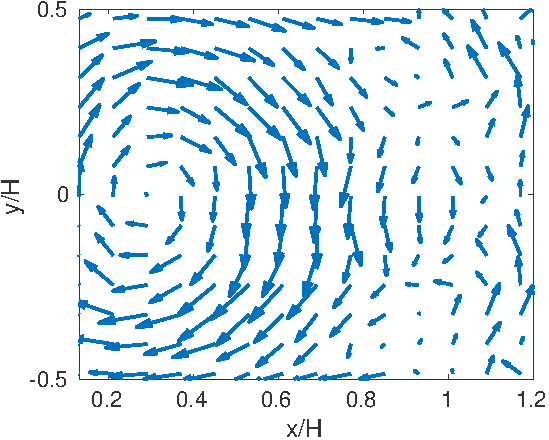} \\ 
\includegraphics[height=32mm]{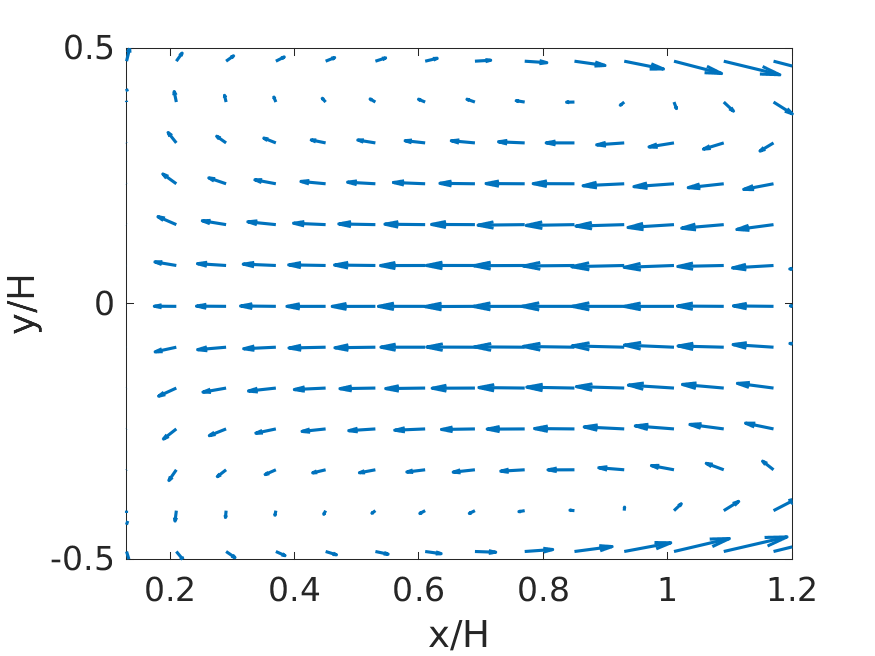} & 
\includegraphics[height=32mm]{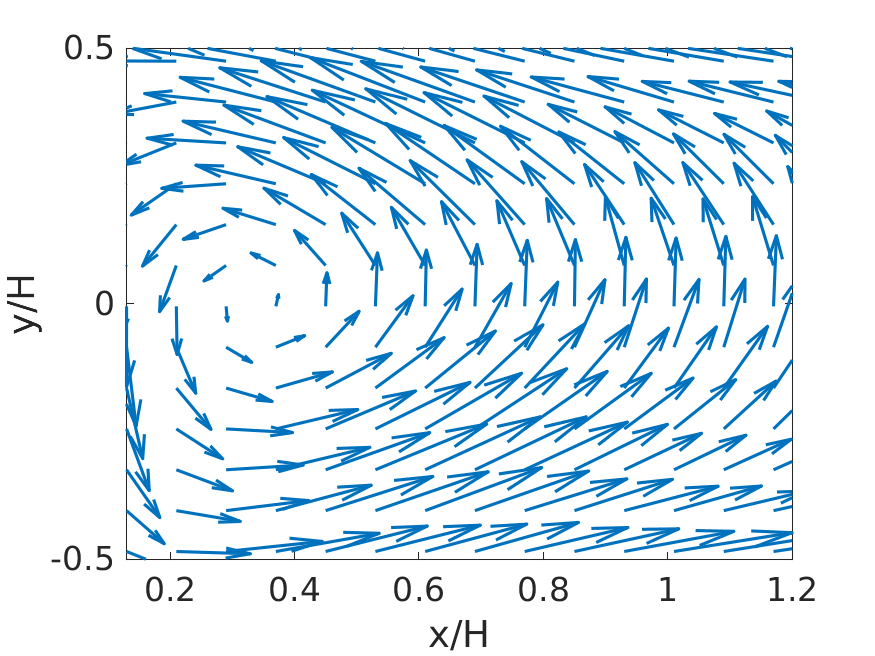} & 
\includegraphics[height=32mm]{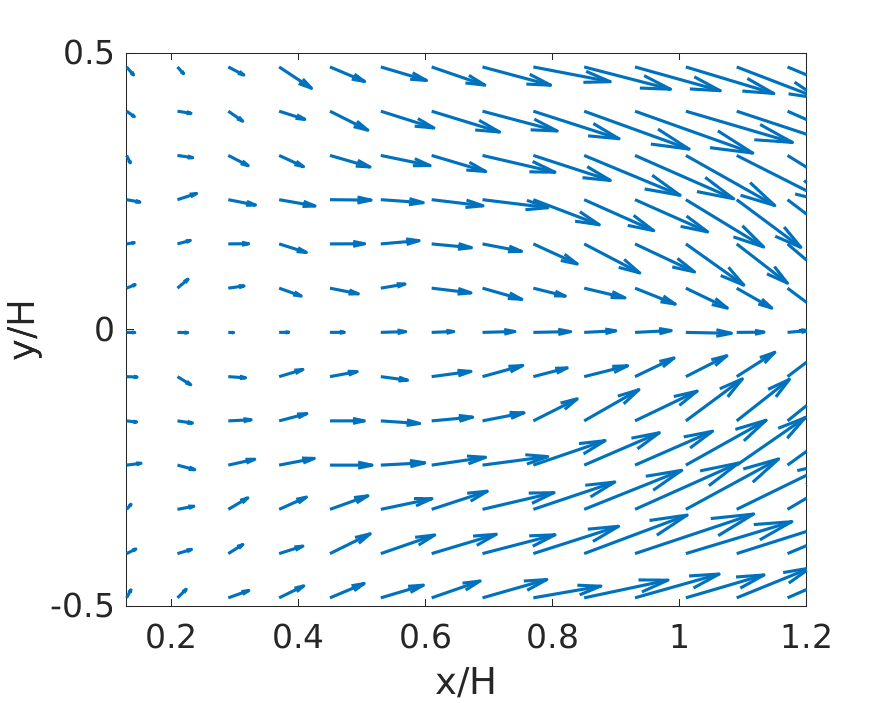} & 
\includegraphics[height=32mm]{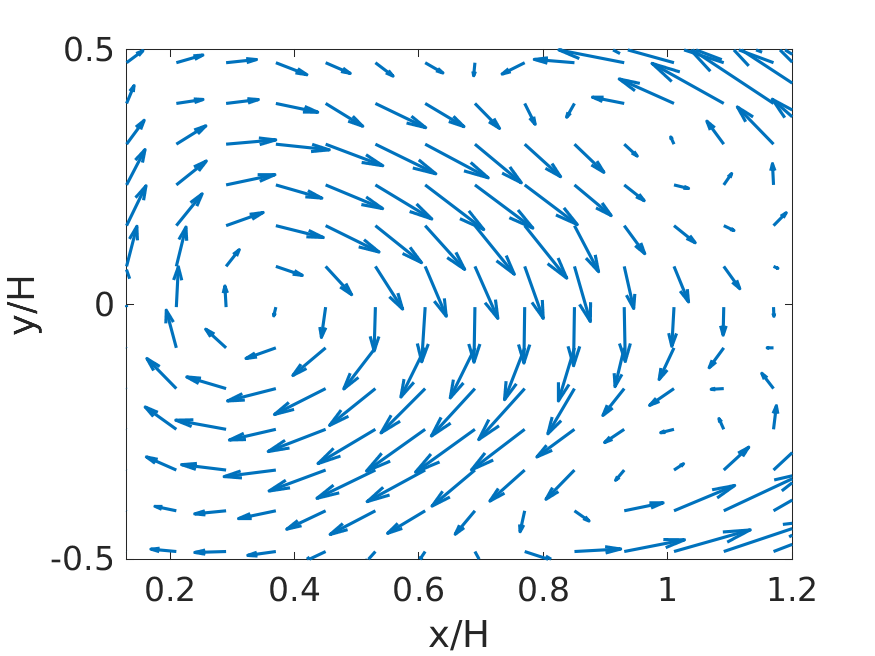} \\ 
\end{tabular}
 \caption{ 2-D wake POD modes 1 to 4; top row: simulation; bottom row: experiment.    } 
 \label{fig4:podmodes14}
\end{figure}

Results for the POD spectrum are shown in figure ~\ref{fig3:lambda2d}
and show a good agreement between the experiment and the simulation.
The second mode represents about 35\% of the total fluctuating horizontal 
kinetic energy in the wake $\sum_{p \ge 2} \lambda_p$
and the first eight modes capture more than 50\%.
Figure ~\ref{fig4:podmodes14}  
compares the first four POD modes for the experiment and the simulation. 
The first mode  represents a cross-section of a toroidal-like
structure constituting the  recirculation bubble.
The second mode corresponds to a single vortical structure located in the recirculation bubble
close to the rear of the body and  sweeping fluid
from one side of the wake to the other. 
It represents the wake deviation. 
Overall a good agreeement is observed between all the modes found in the experiment
and those found in the simulation.

\begin{figure}[h]
\hspace{-4cm}
\begin{tabular}{cc}
\includegraphics[height=50mm]{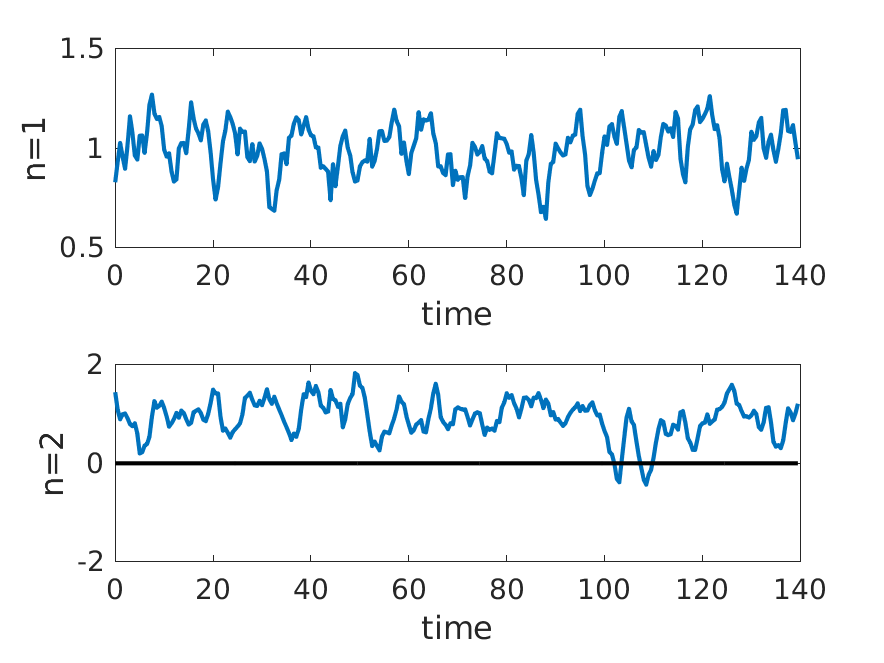} & 
\includegraphics[height=50mm]{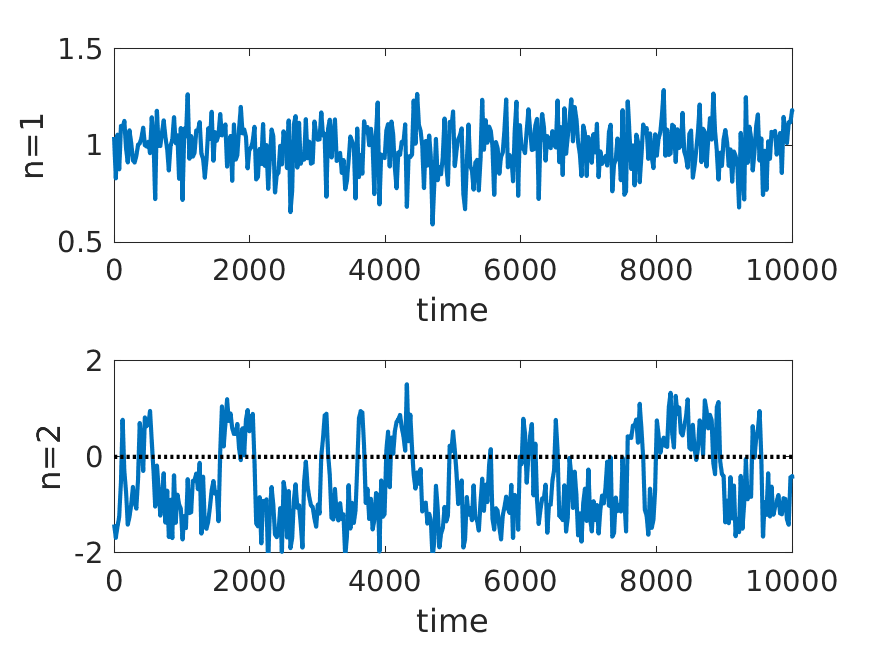} \\ 
\includegraphics[height=50mm]{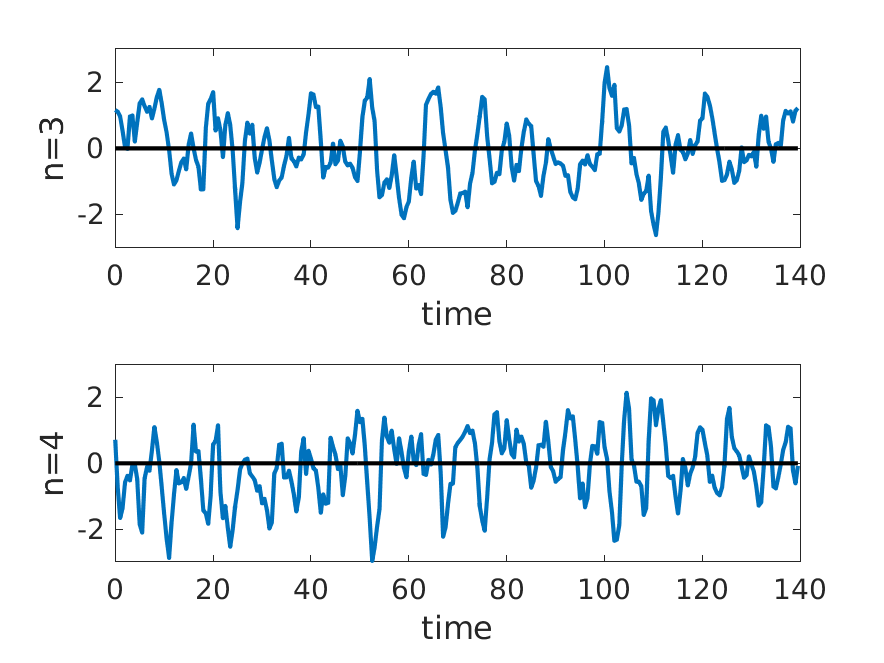} & 
\includegraphics[height=50mm]{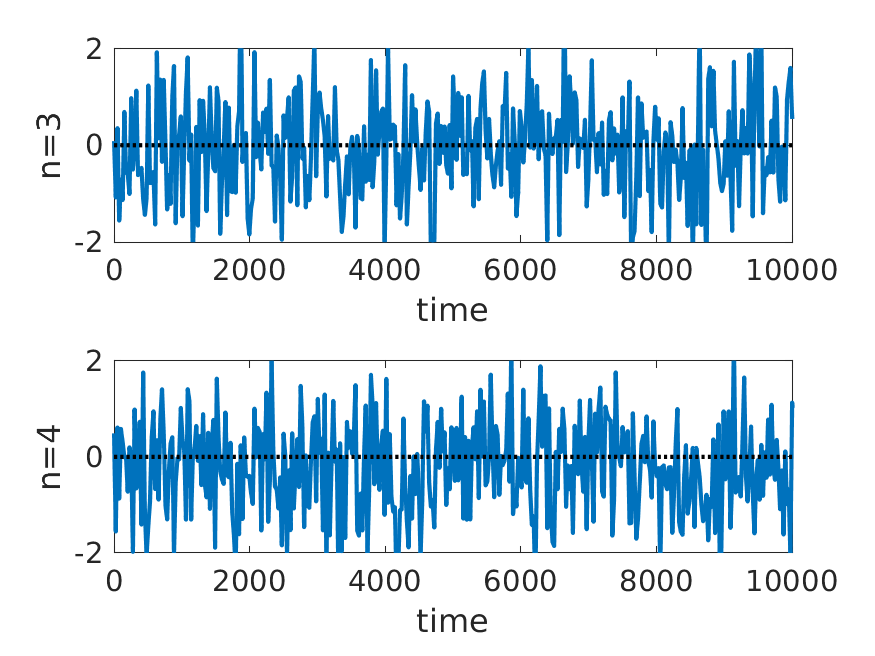} \\ 
\end{tabular}
 \caption{Amplitudes of 2-D POD modes 1  to 4 (from top to bottom); left) 
simulation; right) experiment.}
 \label{fig5:a14_2d}
\end{figure}

Figure ~\ref{fig5:a14_2d} shows for both the experiment and the simulation  the corresponding 
amplitudes (normalized by the square root of the eigenvalue) of the modes 
represented in  figure ~\ref{fig4:podmodes14}.
The amplitudes of the first two modes  in the simulation (figure ~\ref{fig5:a14_2d} right) 
display small oscillations around a constant positive value.
In the experiment (figure ~\ref{fig5:a14_2d} left), the amplitude of the first mode oscillates near a constant positive value, while the amplitude
of the second mode changes sign several times, which corresponds to a switch in the wake asymmetry.

\begin{figure}[h]

\hspace{-1in}
\begin{tabular}{cccc}
\hh n=1 & \hh n=2 & \hh n=3 & \hh n=4 \\
\includegraphics[height=30mm]{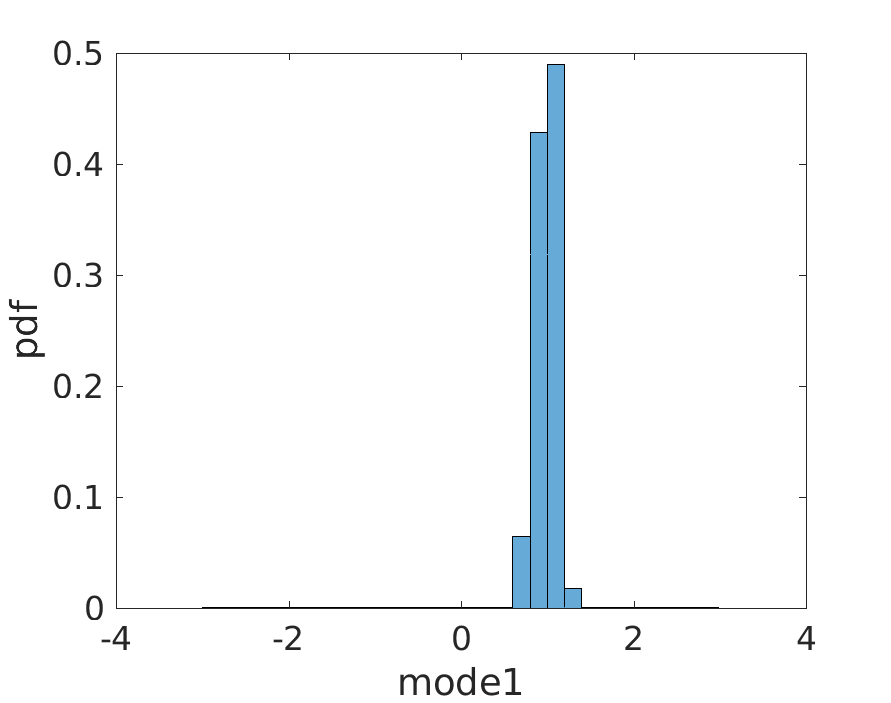} & 
\includegraphics[height=30mm]{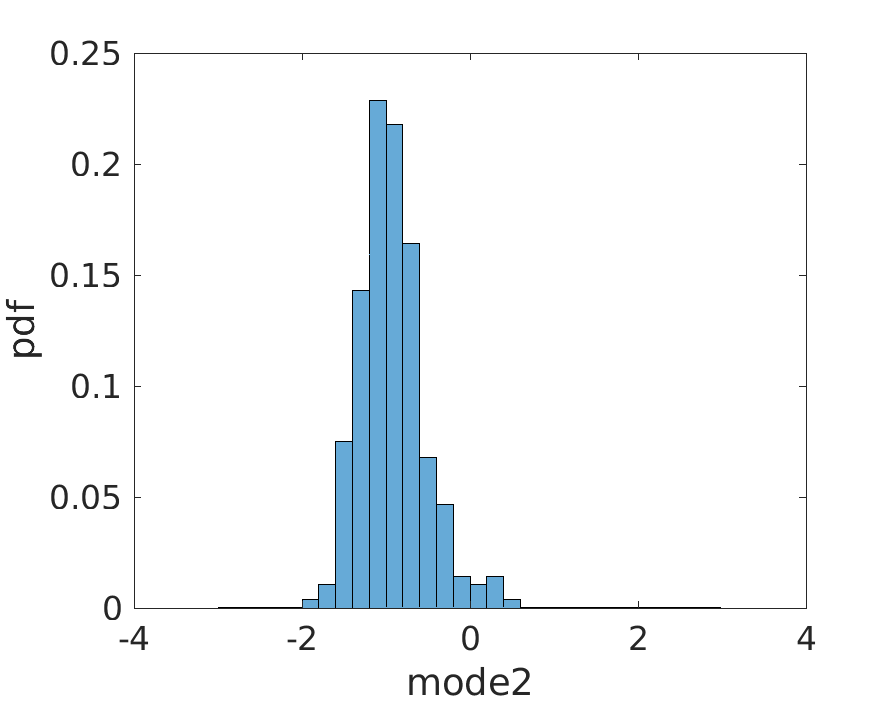} & 
\includegraphics[height=30mm]{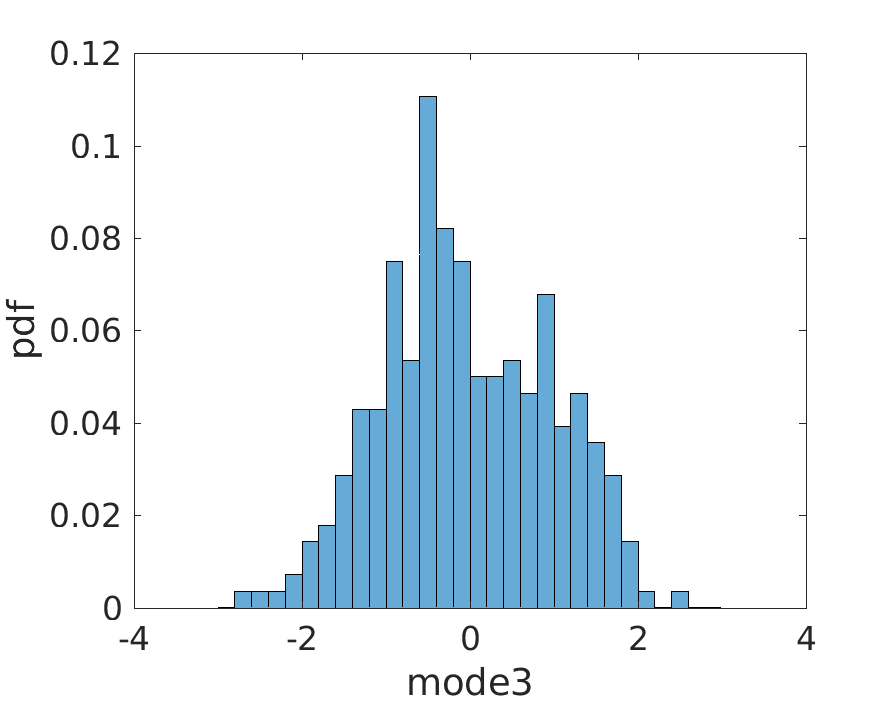} & 
\includegraphics[height=30mm]{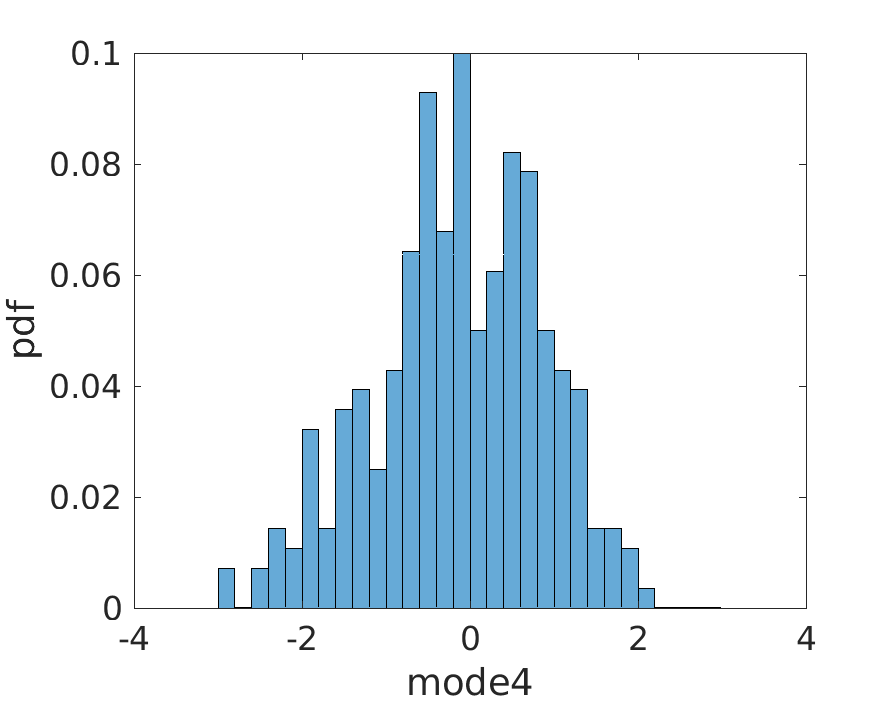} \\ 
\includegraphics[height=30mm]{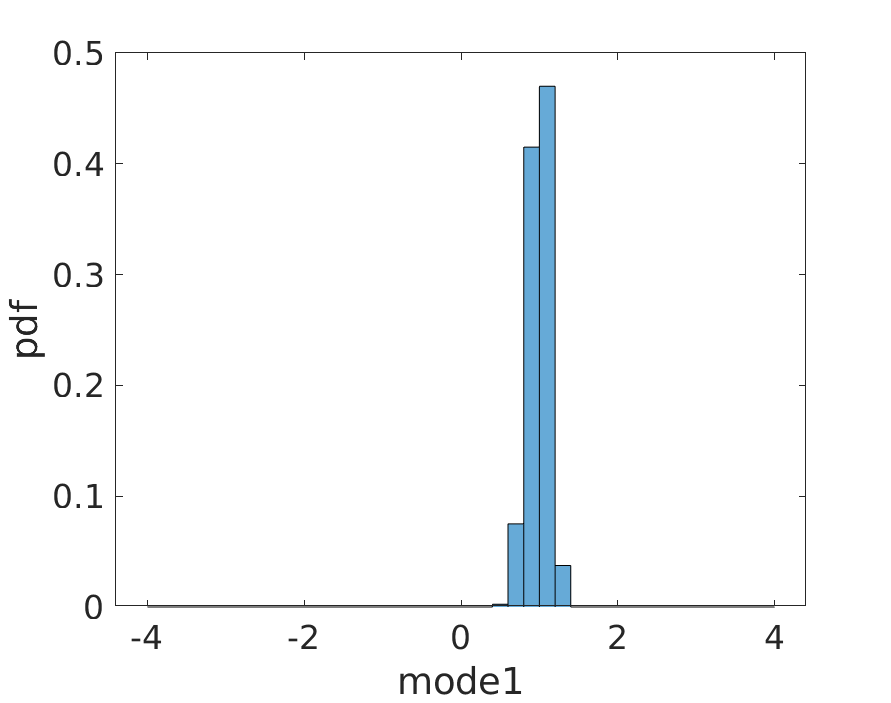} & 
\includegraphics[height=30mm]{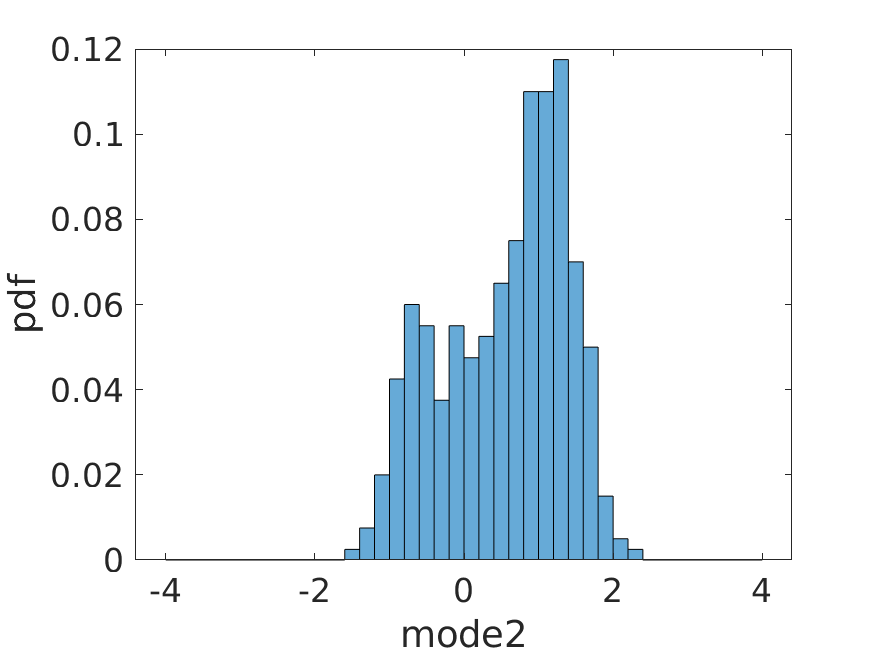} & 
\includegraphics[height=30mm]{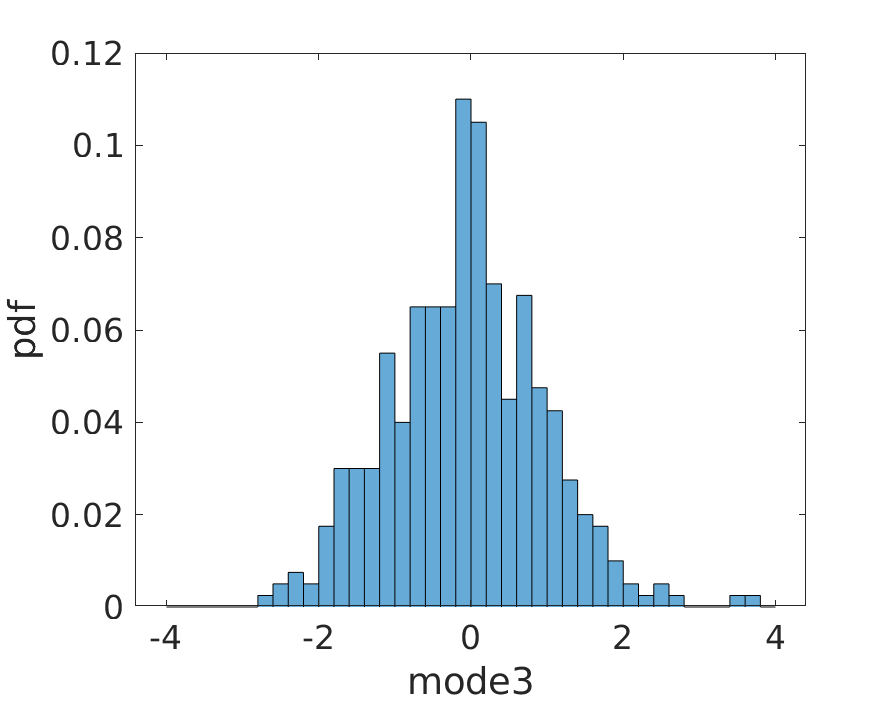} & 
\includegraphics[height=30mm]{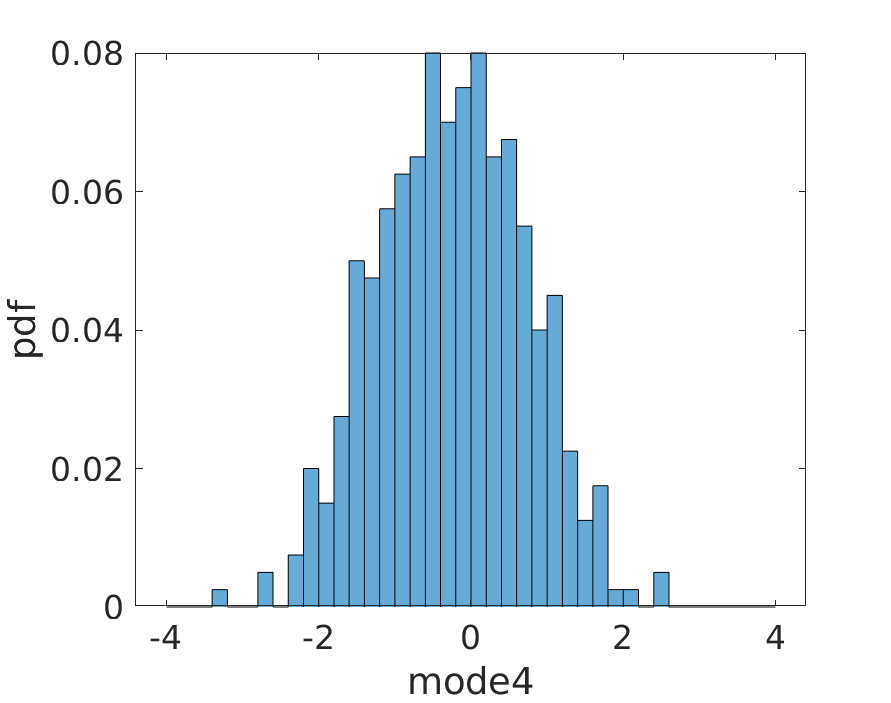} \\ 

\end{tabular}
 \caption{ Histogram comparison of 2-D POD modes 1 to 4 (from left to right);  
top row: simulation; bottom row: experiment.} 
 \label{fig6:hista14}
\end{figure}

We can also see from figure ~\ref{fig4:podmodes14} that 
mode 3 is symmetric and mode 4 is antisymmetric. 
Mode 3  consists of  a longitudinal converging (or diverging, depending on the sign of the amplitude)
motion at the extremity of the recirculation bubble, which could be associated with wake pumping
i.e successive enlargment and shrinking of the recirculation region.
Mode 4   consists of three vortical structures - one larger structure
extending across the wake and located close to the rear of the body, and two smaller ones,
both rotating in the opposite direction, on each side of the recirculating bubble. 
Its action is therefore to distort the recirculation bubble in the spanwise direction. 
Since the PIV results are not resolved in time and the simulation total time is relatively short, it is
difficult to compare directly time evolutions. However histograms
of the amplitudes can be compared in figure ~\ref{fig6:hista14} (we note that only the amplitudes corresponding
to the original set of snapshots are shown).
We can see that there is a relatively good agreement between the experiment
and the simulation.

\begin{figure}[h]
\hspace{-2in}
\begin{tabular}{cccc}
\hh n=5 & \hh n=6 & \hh n=7 & \hh n=8 \\
\includegraphics[height=30mm]{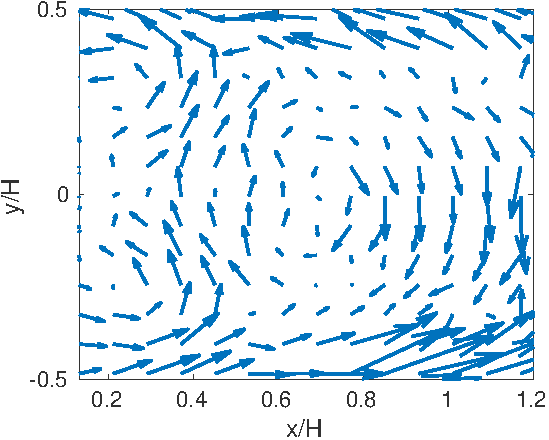} & 
\includegraphics[height=30mm]{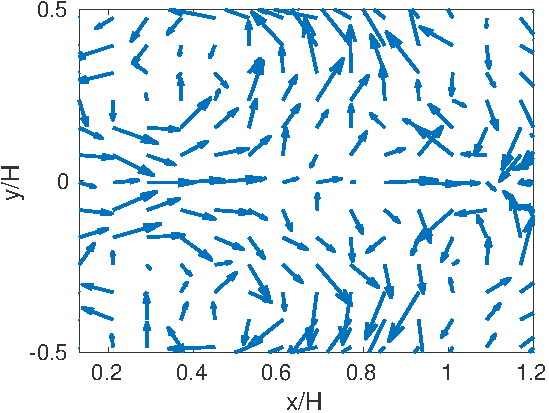} & 
\includegraphics[height=30mm]{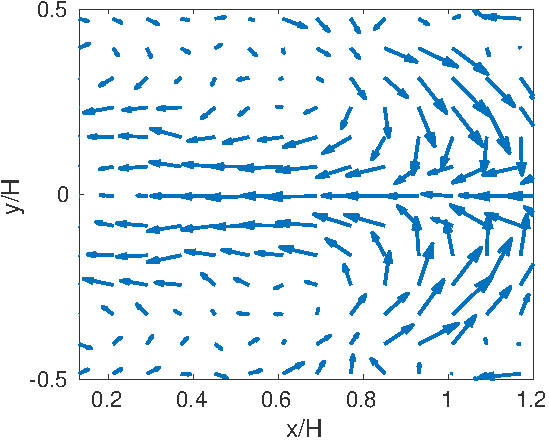} & 
\includegraphics[height=30mm]{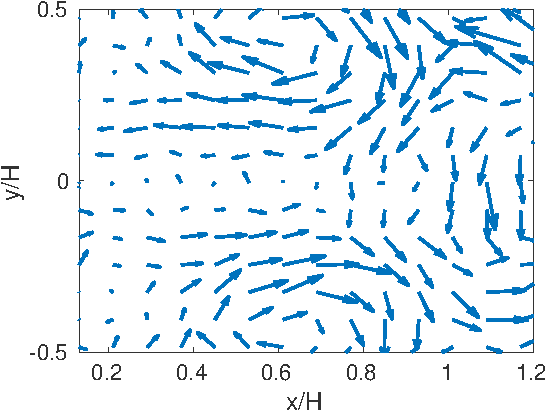} \\ 
\includegraphics[height=32mm]{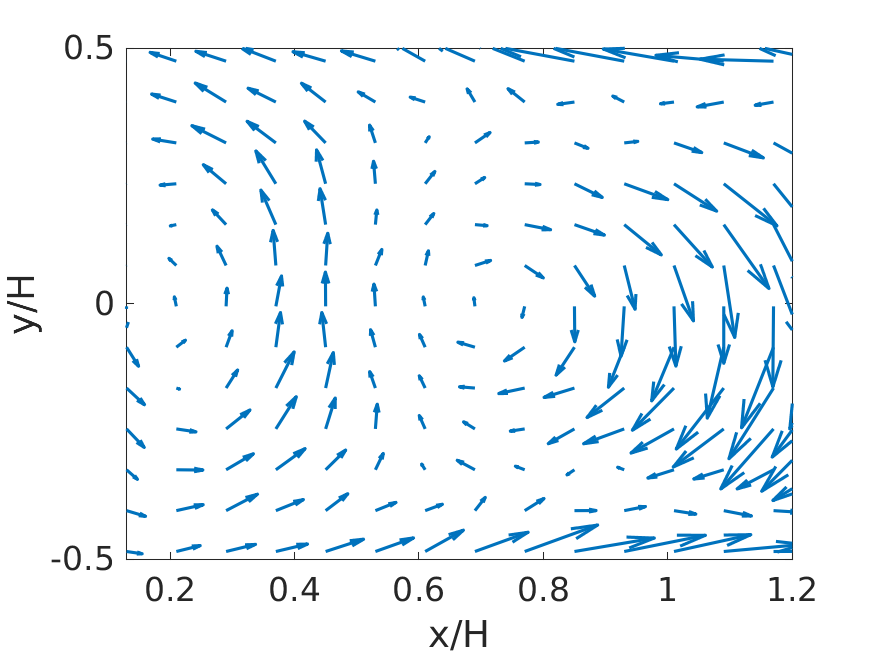} & 
\includegraphics[height=32mm]{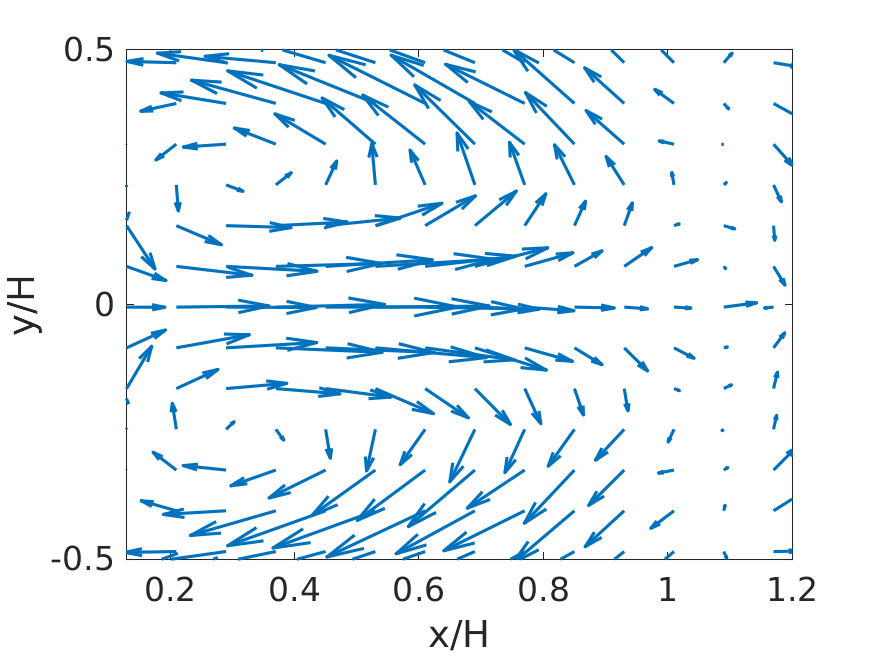} & 
\includegraphics[height=32mm]{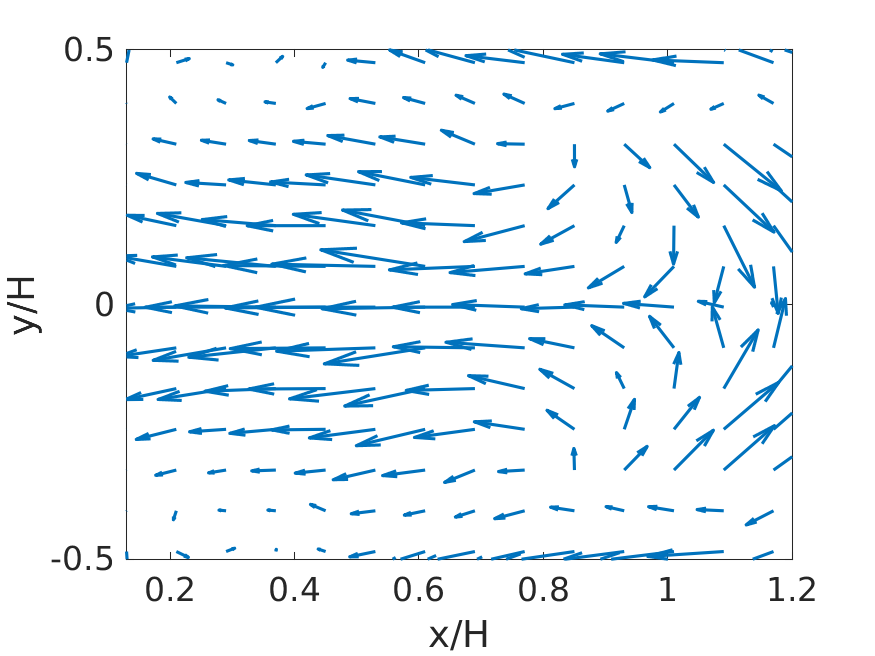} & 
\includegraphics[height=32mm]{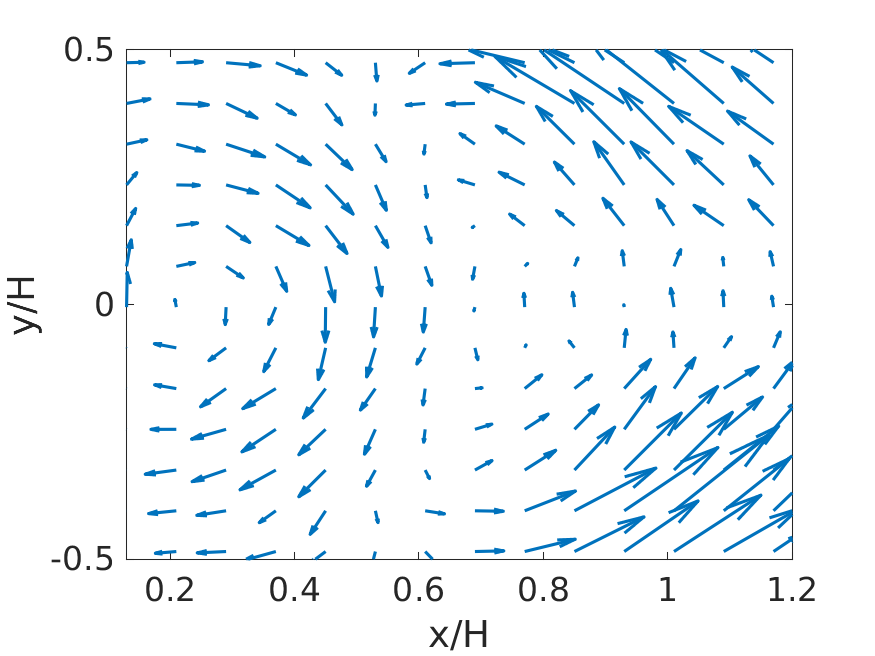} \\ 
\end{tabular}
 \caption{ 2-D wake POD modes 5 to 8; top row: simulation; bottom row: experiment.  } 
 \label{fig7:podmodes58}
\end{figure}

The next four order modes in the simulation, shown 
in figure ~\ref{fig7:podmodes58}, 
also present similarities with the modes in the experiment.
Modes  5 and 8  are antisymmetric and consist of vortical motions 
respectively dominant on the inner part and the outer part
of the recirculation. 
Modes 6 and 7 are symmetric and consist of two  counter-rotating 
vortical structures, extending over the full recirculation length and  located on each side of the wake.
Given the differences in Reynolds number and time resolution, the agreement is 
remarkable and suggests that the most energetic structures have common 
features over a wide range of Reynolds numbers. 

\begin{figure}[h]
\hspace{-3in}
\centerline{
\includegraphics[height=80mm]{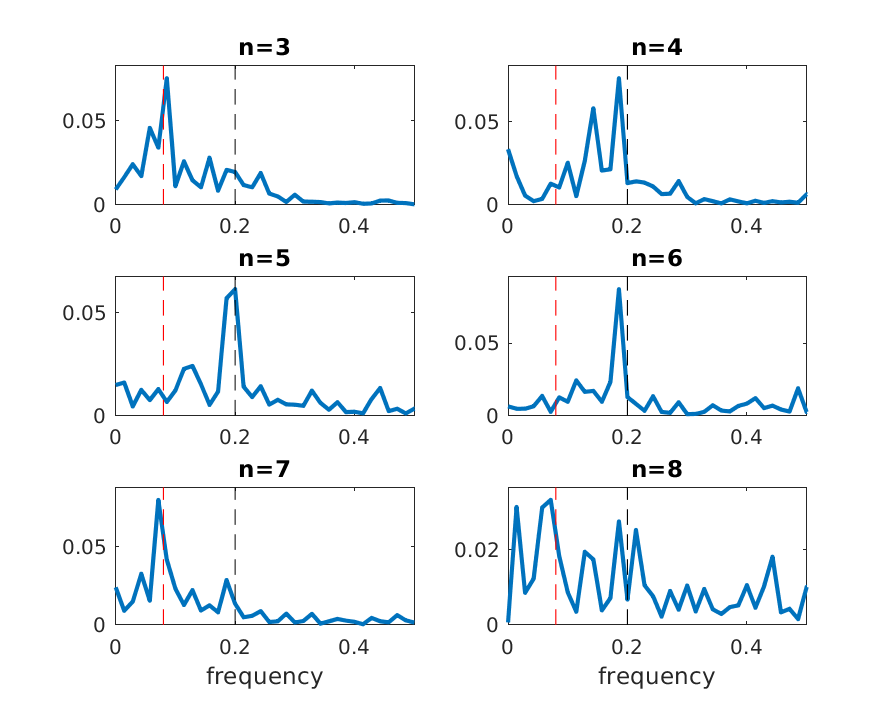} } 
 \caption{ 2-D POD amplitude power spectral density $|\hat{a}_n|^2$
for modes 3 to 8 -
the red and black dashed lines respectively correspond to the frequencies
0.08 and 0.2. } 
 \label{fig8:af_2d}
\end{figure}

The spectral content of   the temporal coefficients is presented in figure ~\ref{fig8:af_2d}. 
All modes are characterized by low frequencies.
The two red and black lines respectively correspond to the frequencies
0.08 and 0.2.  Modes 3  and 7, which are both symmetric, are characterized 
by a dominant frequency around 0.08, while mode 4, 5 and 6, which are antisymmetric, are 
characterized by a frequency of 0.2.  This frequency, which is associated with vortex 
shedding, can be identified most clearly in modes 5 and 6. 
Mode 8 is characterized by a mixture of frequencies.

\section{3-D POD}

\begin{figure}[h]
\hspace{-3in}
\centerline{
\includegraphics[height=70mm]{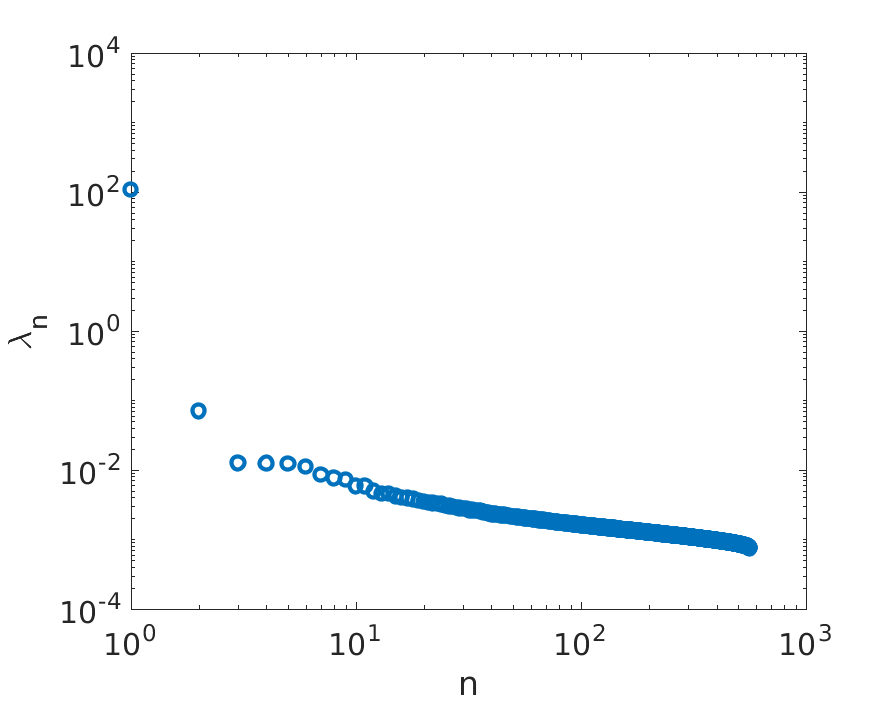} } 
\caption{3-D POD eigenvalue spectrum.}
\label{fig9:lambda3d}
\end{figure}

\begin{figure}
\hspace{-3in}
\includegraphics[height=60mm]{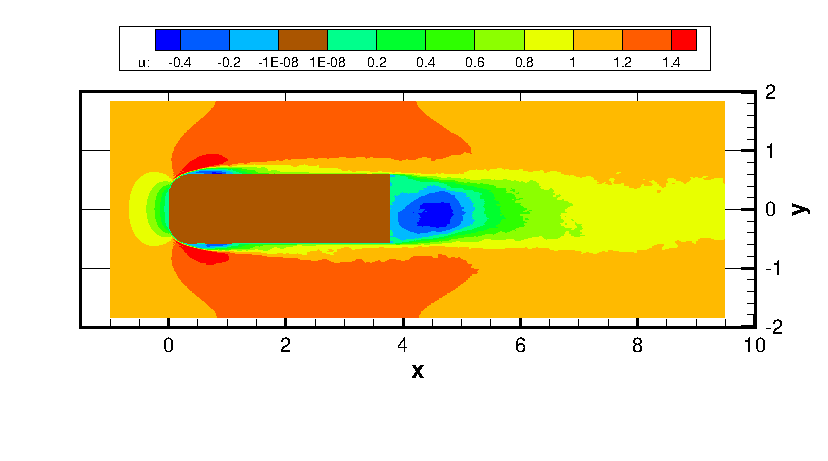}   \\
\hspace{-3in}
\includegraphics[height=60mm]{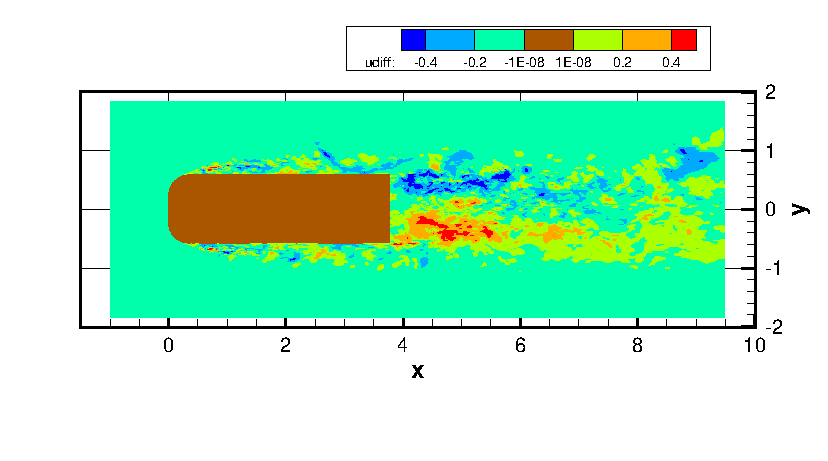}  
\caption{Streamwise velocity contours on mid-height 
plane $z=-0.5$; top : projection of an instantaneous 
field $\uu$  onto the first ten modes $u_{recons} = \sum_{n=1}^{10} a_n(t) \pphi(\xx)$ where
$a_n=\int \uu(\xx,t) . \pphi(\xx) d\xx.$; bottom: difference field 
$| u  - u_{recons}| $.} 
\label{fig10:recons10}
\end{figure}

To investigate the flow structure and dynamics in more detail,
POD is applied to the full 3-D velocity field over the entire computational domain.
The eigenvalue spectrum is shown in figure ~\ref{fig9:lambda3d}.
The first 3-D mode, which coincides with the mean flow, 
is much more energetic than the other modes 
compared with 2-D analysis (figure ~\ref{fig3:lambda2d}) 
since the entire domain contains a large steady contribution
of the kinetic energy outside the wake.
Due to the extent of the spatial domain, the convergence of the fluctuations 
is much slower than in the 2-D wake measurements.  
The second mode represents 
only 8\% of the total fluctuating kinetic energy  $\sum_{p \ge 2} \lambda_p$, 
\berengere{which is equivalent to the combined energy of the next 
eight most energetic modes. }  
We note that the modes 3 to 6 have nearly similar eigenvalues.
This reflects the fact that the physical structures are global, three-component modes defined over the full domain, linking
the upstream flow, the four boundary layers along the body, the near and the far wake. 
Figure ~\ref{fig10:recons10} shows the projection  of an instantaneous velocity field 
on the first ten modes and the difference between the full field 
and its projection.
It is clear that only the large scales are captured by the projection.  
One can see that most of the unresolved modes (which represent the major portion of the total
kinetic energy) are located in the boundary layers, the shear layers and the far wake.

\begin{figure}[h]
\hspace{-2in}
\begin{tabular}{ccc}
n=1 & \includegraphics[align=c,height=28mm]{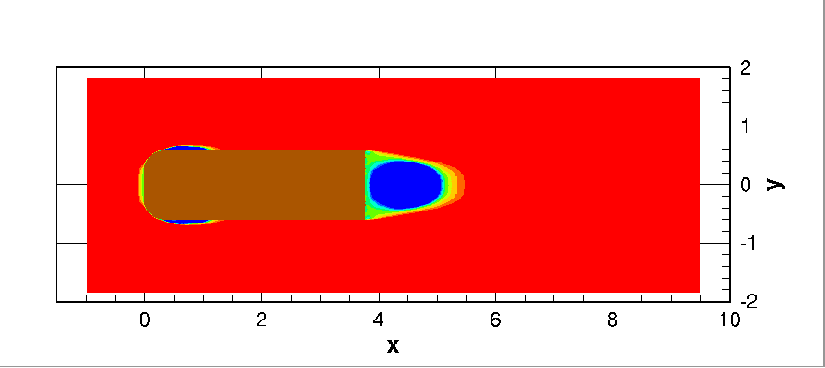}
& \includegraphics[align=c,height=28mm]{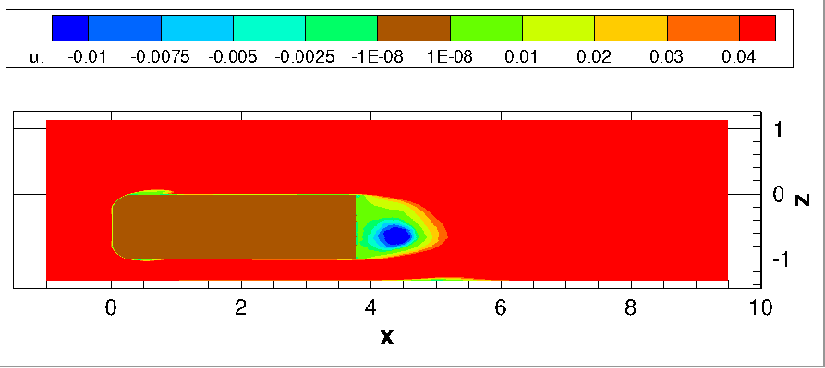} \\
n=2 &  \includegraphics[align=c,height=28mm]{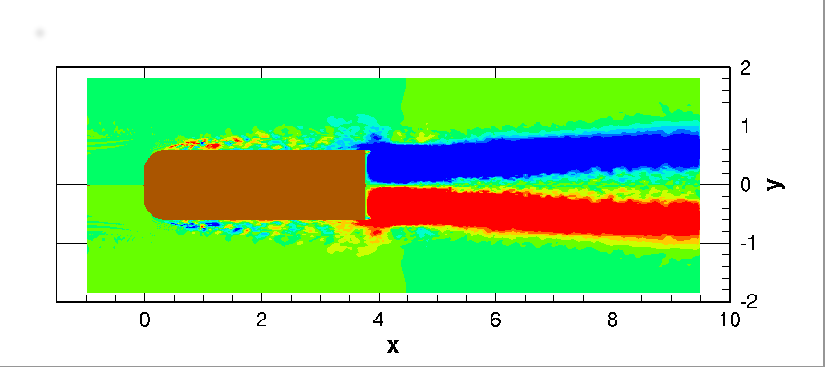}&
\includegraphics[align=c,height=28mm]{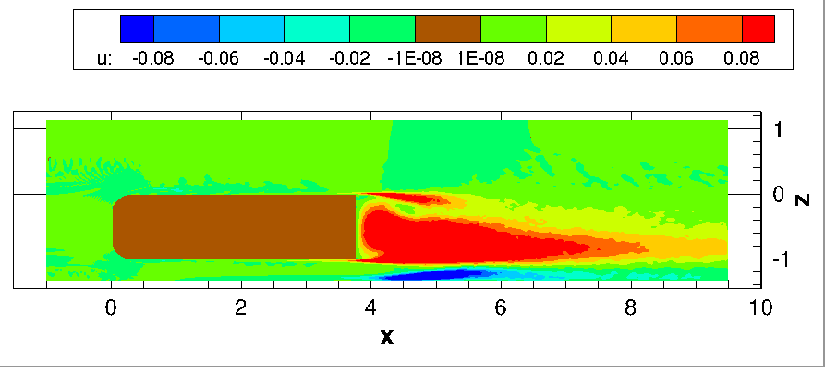} \\

\end{tabular}
 \caption{ Streamwise velocity contours of 3-D POD modes 1 (top row)
and 2 (bottom row); left) horizontal section 
at mid-height $z=-0.5$; right) vertical section at $y=-0.4$.} 
 \label{fig11:modes12}
\end{figure}

We now investigate the 
properties of the POD modes.
For spatial characteristics, we represent
in figures ~\ref{fig11:modes12} and ~\ref{fig14:modes310} the streamwise velocity contour of the 3-D POD modes in both  a horizontal cross-section at mid-height $z=-0.5$, and in a vertical plane located in an off-center spanwise location $y=-0.4$ (since asymmetric motions will cancel on the symmetry plane).
The first two spatial modes are shown in figure ~\ref{fig11:modes12}.
Although the representation of the first two POD modes
in 2-D and 3-D is different (figures ~\ref{fig4:podmodes14} and ~\ref{fig11:modes12}), 
it can be shown that the restriction of the  first two 3-D POD 
modes to the cross-section of the wake is similar to 
the 2-D POD results, which 
is not entirely surprising since they are expected to represent 
the symmetric and antisymmetric part of the time-averaged velocity field. 

\begin{figure}
\hspace{-3in}
\begin{tabular}{cc}
\includegraphics[height=60mm]{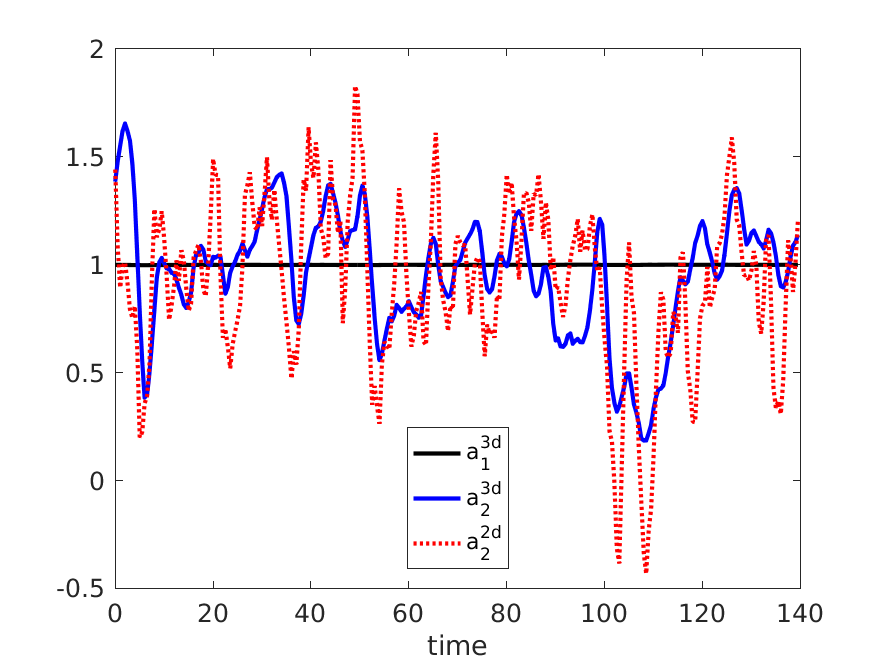} &
\includegraphics[height=60mm]{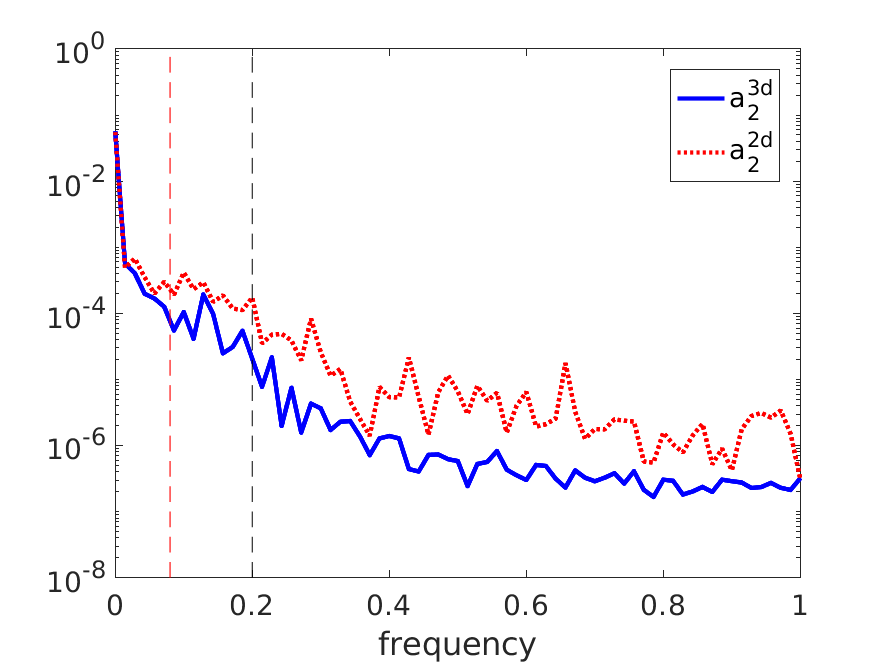} \\
\end{tabular}

\caption{
 Left)  amplitudes of the first two POD modes;
right) power spectral density of the quasi-steady deviation mode
$|\hat{a}_2|^2$ -
the red and black lines respectively correspond to the two 
frequencies 0.08  and 0.2.} 
\label{fig12:af3d12}
\end{figure}

\begin{figure}
\hspace{-2in}
\begin{tabular}{cc}
\includegraphics[height=65mm]{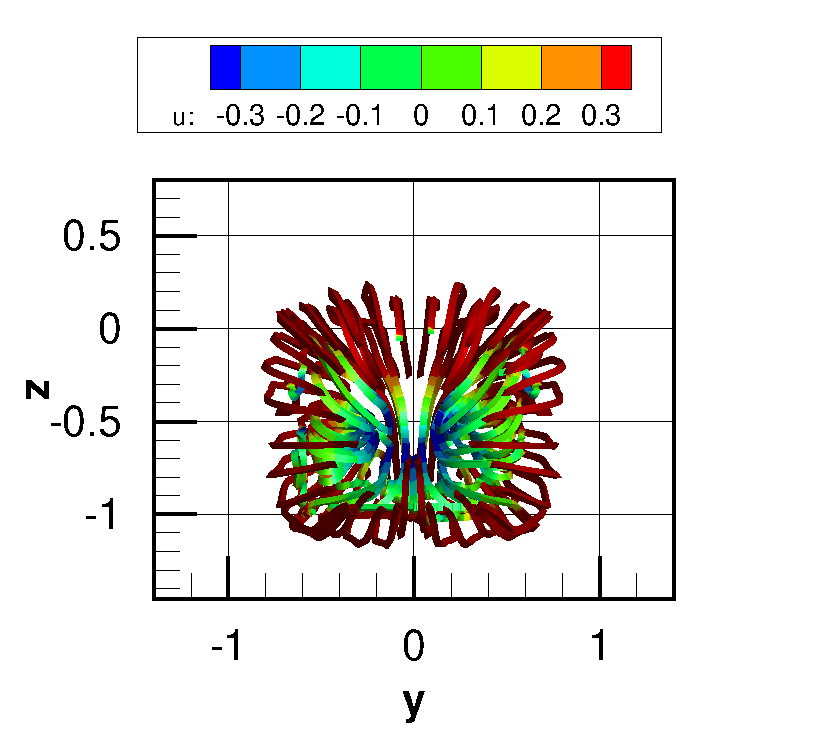}  &
\includegraphics[height=65mm]{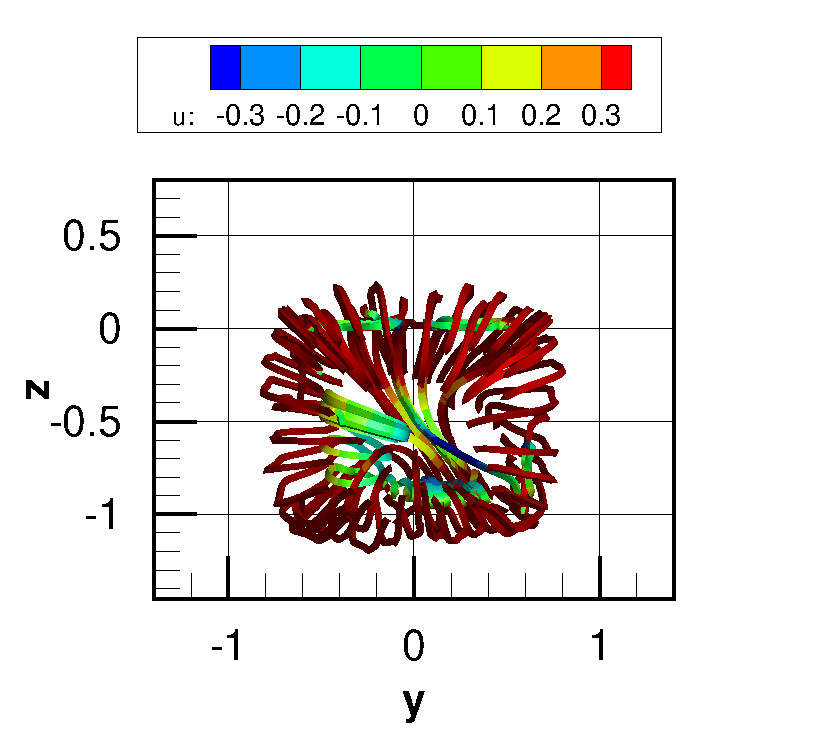}  \\
\end{tabular}
\caption{Streamlines of the field viewed from downstream  
left) mode 1 $\sqrt{\lambda_1} \pphi_1$; right) combination of mode 1 and mode 2
$\sqrt{\lambda_1} \pphi_1 + \sqrt{\lambda_2} \pphi_2$ .}
 \label{fig13:stream2}
\end{figure}

\begin{figure}[h]
\begin{tabular}{cccc}
& \includegraphics[height=6mm]{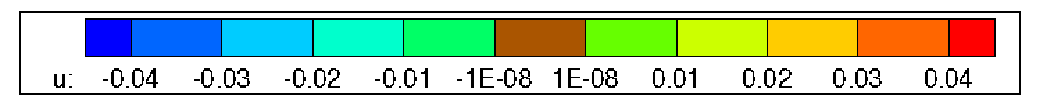}  & \\
mode &  \hspace{2cm} u contour at $z=-0.5$ & 
 \hspace{2cm} u contour at $y=-0.4$ \\
\raisebox{14mm}{n=3} & 
\includegraphics[height=23mm]{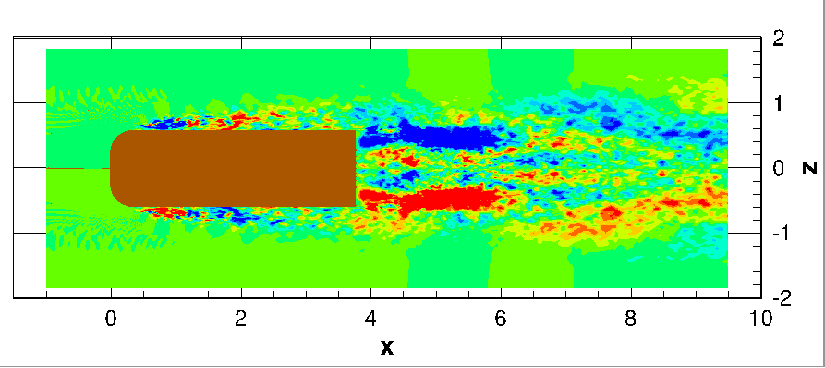}  &
\includegraphics[height=23mm]{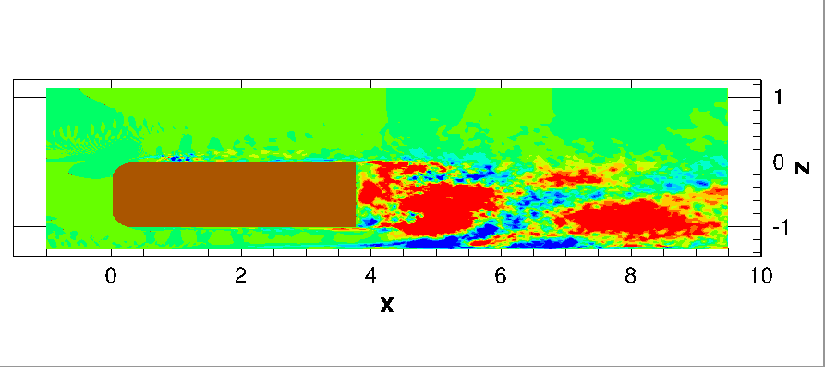}  \\ 
\raisebox{14mm}{n=4} & 
\includegraphics[height=23mm]{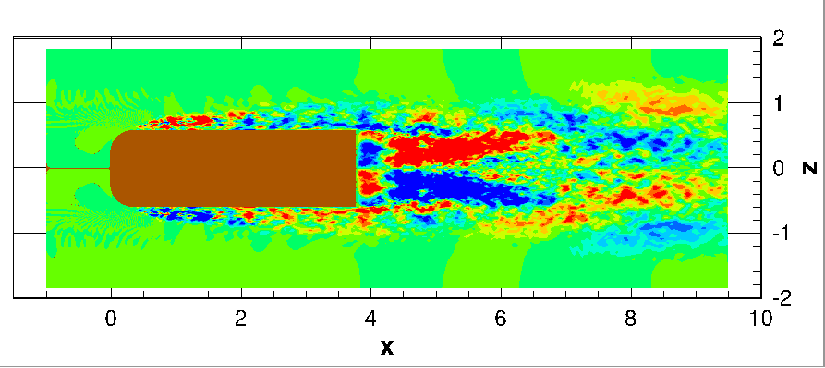}  &
\includegraphics[height=23mm]{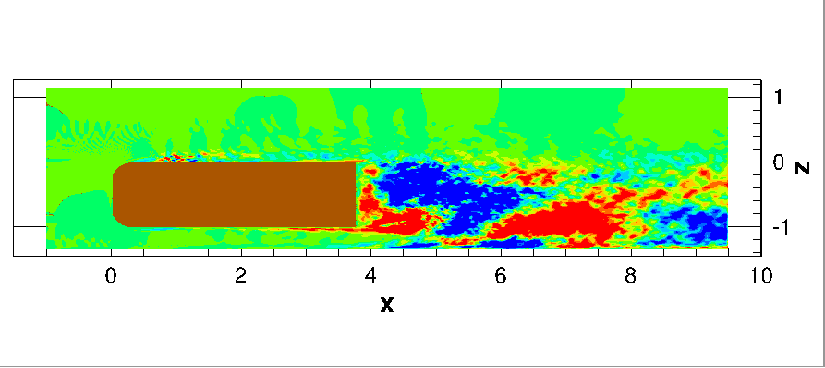}  \\ 
\raisebox{14mm}{n=5} & 
\includegraphics[ height=23mm]{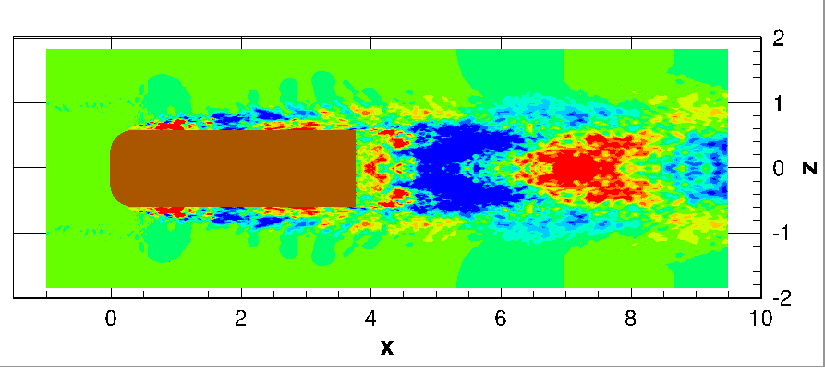}  &
\includegraphics[height=23mm]{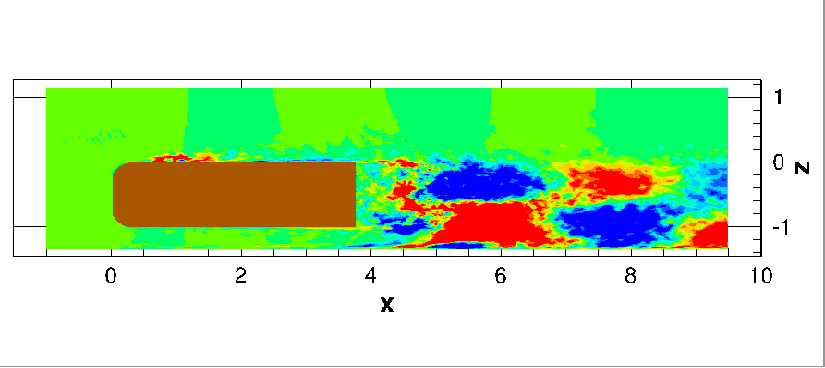}  \\ 
\raisebox{14mm}{n=6} & 
\includegraphics[height=23mm]{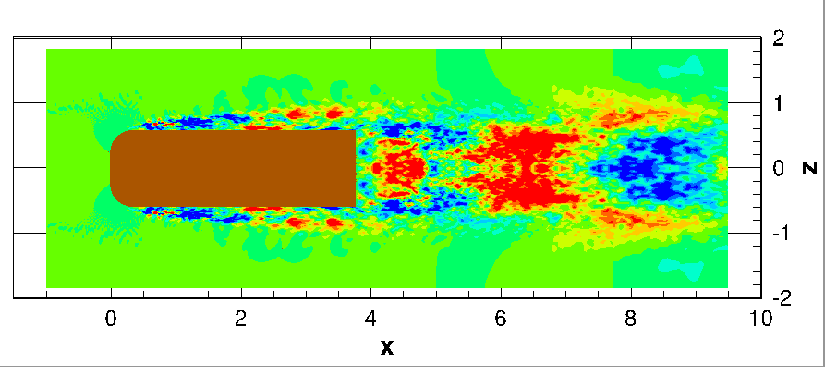}  &
\includegraphics[height=23mm]{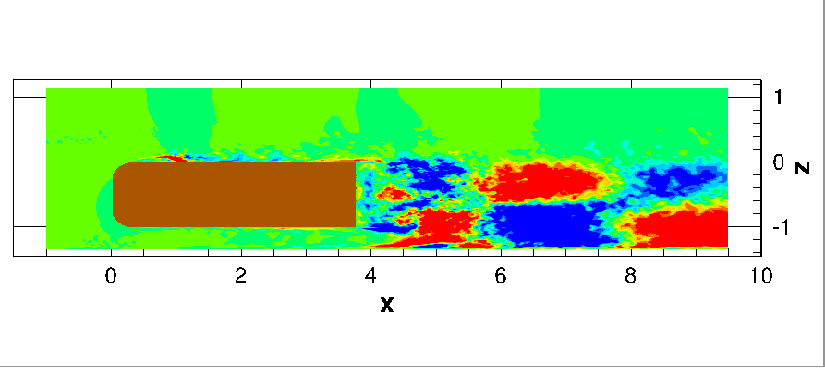}  \\ 
\raisebox{14mm}{n=7} & 
\includegraphics[height=23mm]{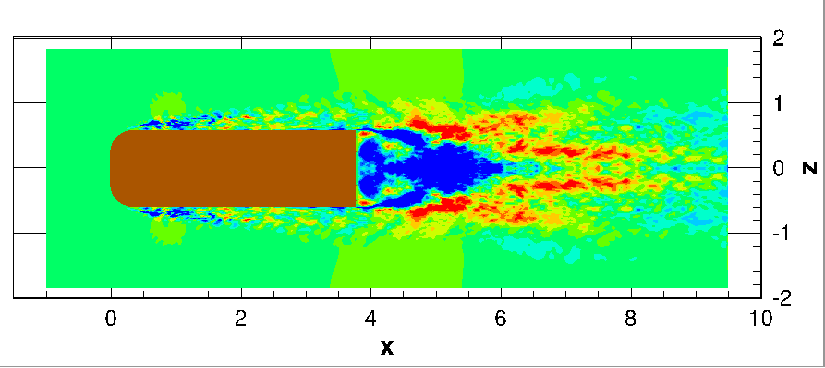}  &
\includegraphics[height=23mm]{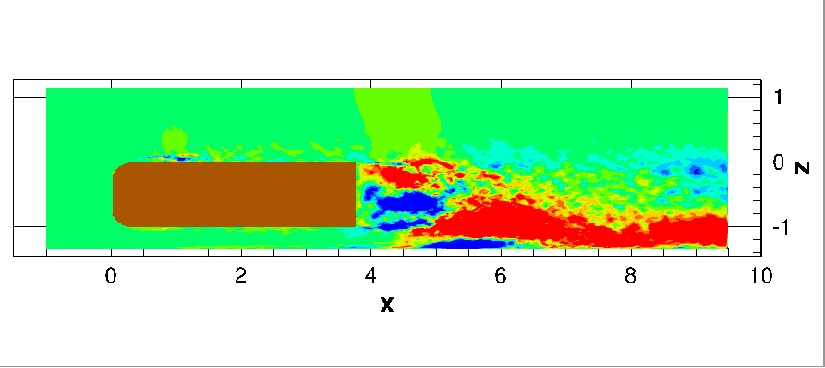}  \\ 
\raisebox{14mm}{n=8} & 
\includegraphics[height=23mm]{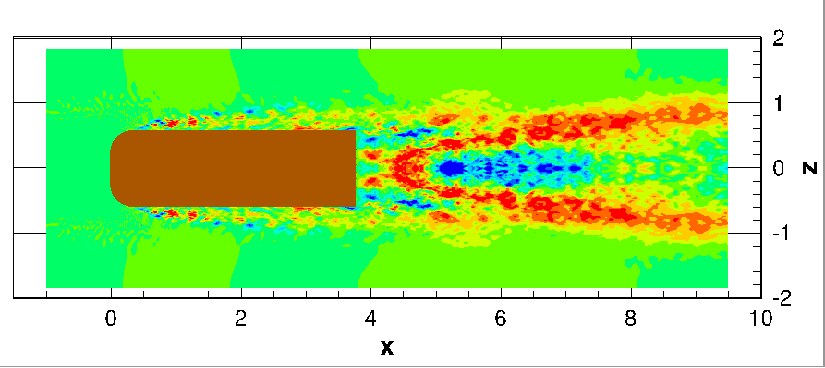}  &
\includegraphics[height=23mm]{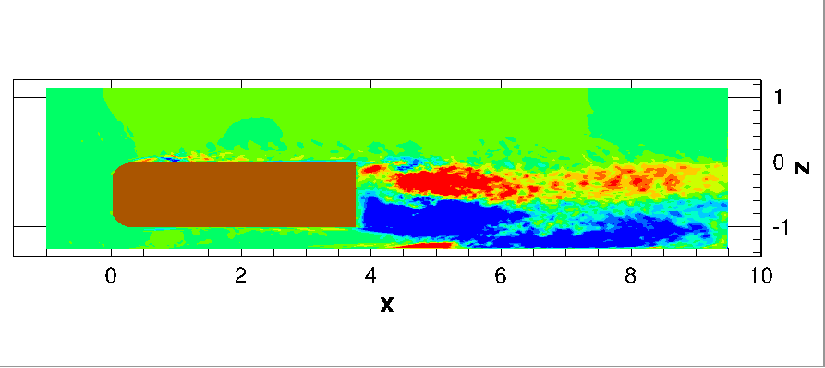}  \\ 
\raisebox{14mm}{n=9} & 
\includegraphics[height=23mm]{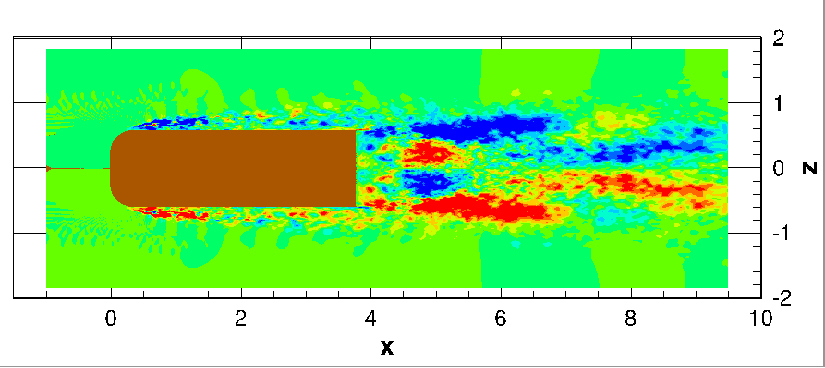}  &
\includegraphics[height=23mm]{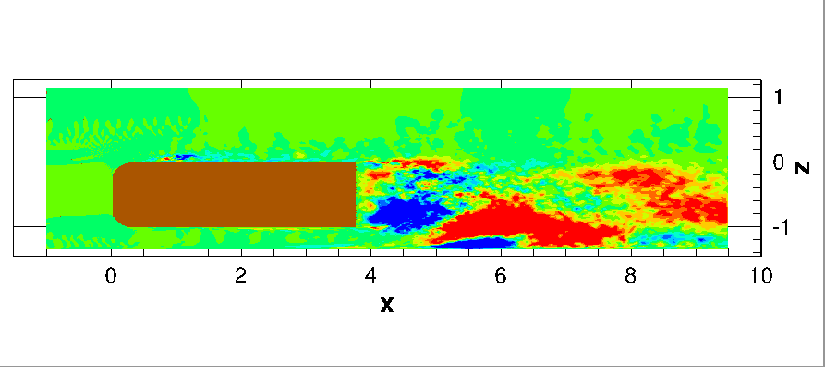}  \\ 
\raisebox{14mm}{n=10} & 
\includegraphics[height=23mm]{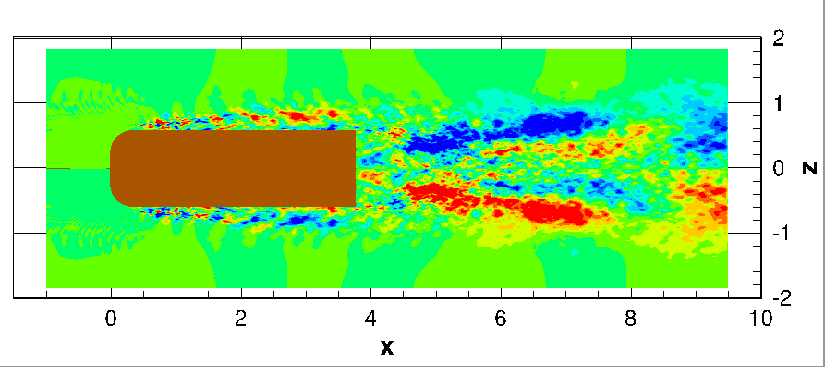}  &
\includegraphics[height=23mm]{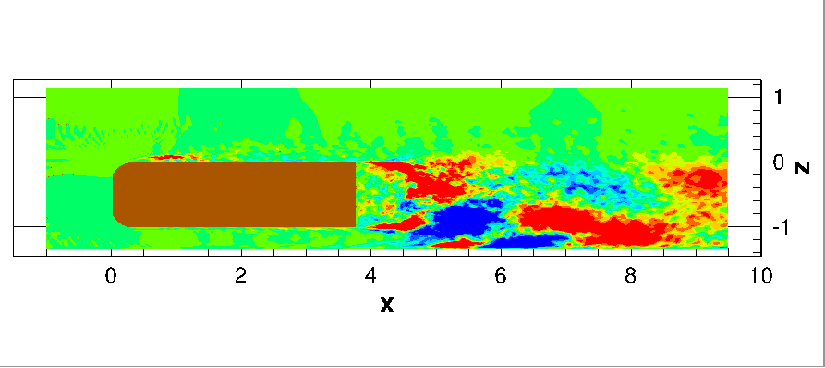}  \\ 
\end{tabular}
 \caption{  Streamwise velocity contours 
of 3-D POD spatial modes 3 to 10 (from top to bottom); 
 left) horizontal section at mid-height $y=-0.5$; right) 
 vertical section at $z=-0.4$.}
 \label{fig14:modes310}
\end{figure}

One can see in figure ~\ref{fig12:af3d12}  that the temporal evolution of the first two 3-D POD modes is slightly different from that of the 2-D POD coefficients.  Unlike its 2-D counter part 
(figure ~\ref{fig5:a14_2d} right), the amplitude of the first 3-D POD mode is essentially constant, which 
shows the statistical convergence of the database. 
Moreover, the amplitudes of the second 2-D and 3-D mode present some discrepancies, but 
the correlation between the mean deviation coefficient $a_2^{2d}$ and $a_{2}^{3d}$ is 0.6,
which indicates that the 2-D measurements are able to describe reasonably well the evolution of 
the 3-D deviation mode.
However, although the amplitude of 2-D POD mode 2 changes sign, the corresponding 3-D POD amplitude $a_2^{3d}$ always remains positive. This means that there are no global switches in the total wake deviation although a local planar measure may indicate otherwise.
In both cases, their corresponding spectrum indicates fluctuations at low frequencies thus confirming their very long-time dynamics evolution. In addition to its permanent asymmetry, the POD mode 2 corresponds to the very low frequency global mode as reported in \cite{kn:rigas15_a,kn:rigas15_b} for an axisymmetric body and responsible for the bistable dynamics found in \cite{kn:grandemange14} for the Ahmed body. 
 To get a better insight into the action of the global deviation mode, figure ~\ref{fig13:stream2} compares 
streamlines in the recirculation zone for the symmetrized mean field (mode 1) and the mean field (modes 1 and 2).
We can see that the effect of the global deviation is to gather the streamlines around one of the base diagonals.

\begin{figure}
\hspace{-2in}
\begin{tabular}{cc}

\includegraphics[height=60mm]{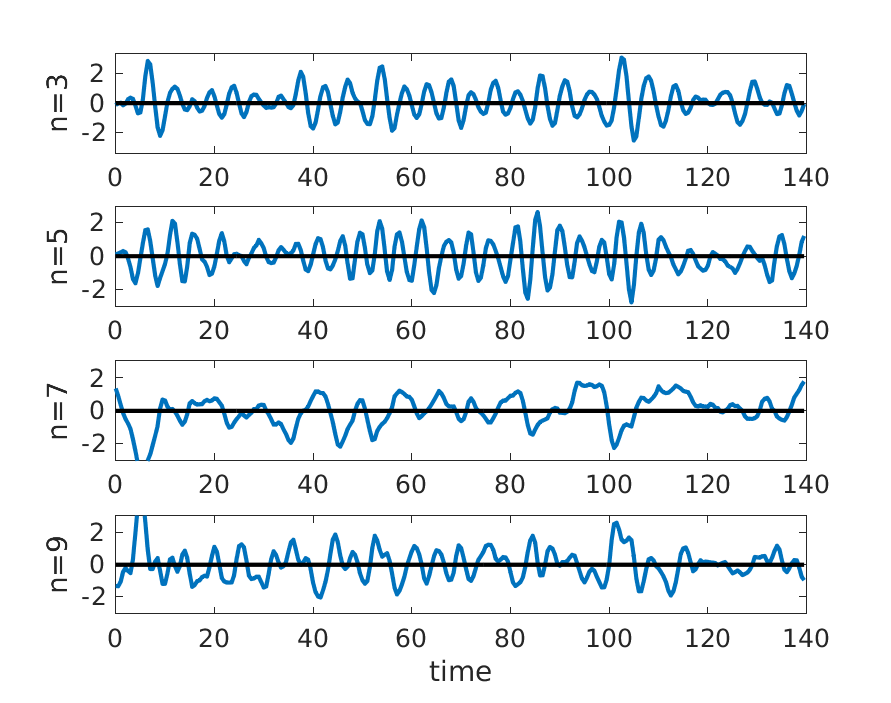} &
\includegraphics[height=60mm]{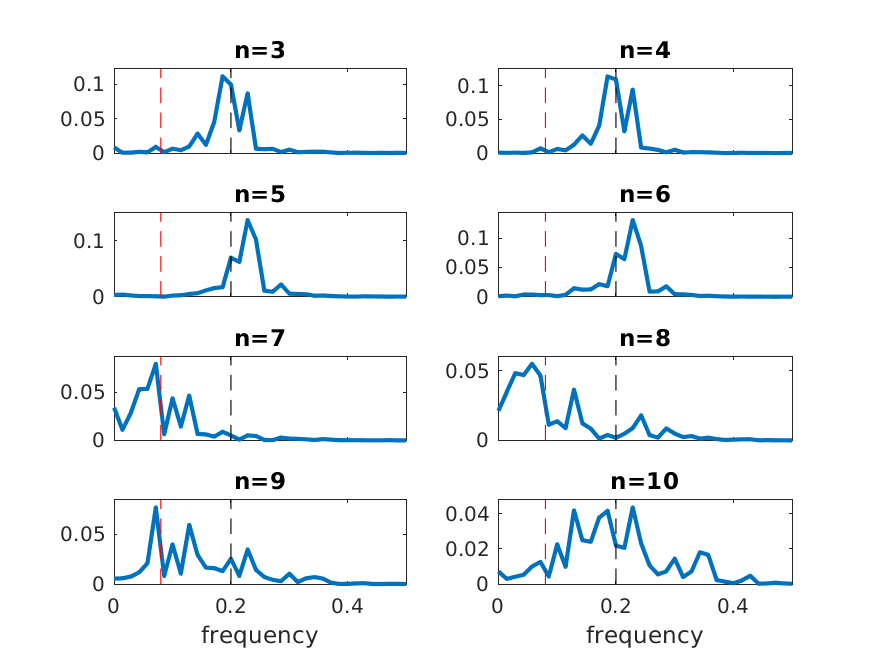} 
\end{tabular}
\caption{
Left) amplitudes of the modes $a_n$, n=3, 5, 7, 9; 
right) power spectral density of the 3-D POD mode amplitudes in the simulation
$|\hat{a}_n|^2$ - 
the red and black lines respectively correspond to the two 
frequencies 0.08  and 0.2.} 
\label{fig15:af3d}
\end{figure}

Modes 3 and 4 are antisymmetric modes which display
a periodicity in the streamwise direction (figure ~\ref{fig14:modes310}). 
The time evolution of these modes is quasi-periodic as shown by the time series and the spectra in figure~\ref{fig15:af3d}. The characteristics frequencies lie in the range 0.19-0.24, with a maximum around 0.19. This frequency matches satisfactorily the \berengere{K\'arm\'an} global mode observed for oscillations in the spanwise direction in (Grandemange {\it et al.} \cite{kn:grandemange14}, Volpe {\it et al.} \cite{kn:volpe15}). Similar observations are made for the symmetric modes 5 and 6 in figure ~\ref{fig14:modes310} and the corresponding temporal properties in figure~\ref{fig15:af3d}. Their spectra display a maximum around 0.23, that is ascribed to the vortex shedding (i.e. \berengere{K\'arm\'an} global mode) with oscillations in the vertical direction. The highest frequency corresponding to the small height and large frequency to the wider width of the body are in agreement with Grandemange {\it et al.} \cite{kn:grandemange14}, Volpe {\it et al.} \cite{kn:volpe15}. This is a general result
for three-dimensional geometries (Kiya and Abe \cite{kn:kiya99}).   

The modes 7 and 8 in figure~\ref{fig14:modes310}  are symmetric
and clearly do not display streamwise periodicity. 
 The action of mode 7 in the horizontal mid-plane is 
to modulate the bubble zone which is either inflated or shrunk, depending
on the sign of $a_7$ (figure ~\ref{fig15:af3d} left). 
When $a_7 >0$,  figure ~\ref{fig14:modes310} shows that 
strong negative fluctuations are present in the bubble, which delays reattachment, and vice-versa. 
The spectra shown in figure ~\ref{fig15:af3d} (right) show that
both modes 7 and 8 are characterized by a strong energetic content
at a frequency of 0.08, which suggests that these modes contribute
significantly to the wake pumping, 
as characterized by Rigas {\it et al.} \cite{kn:rigas14},  Volpe {\it et al.} 
 \cite{kn:volpe15} and Pavia {\it et al.} \cite{kn:pavia18}.

 Modes 9 and 10, which are  
antisymmetric (figure ~\ref{fig14:modes310}), 
are more difficult to interpret.
As noted earlier, there is no guarantee that an individual POD mode 
corresponds to a well-defined physical mechanism.
The spectra in figure~\ref{fig15:af3d} (right)
show that modes 9 and 10 are characterized by a mixture of frequencies in 
an intermediate range 0.08-0.2, with a peak for both modes around 0.15. 
It is therefore likely that these modes correspond to a superposition
of different physical processes. 

As a summary, we find a clear correspondance between the POD modes and the main global modes that contribute to the wake dynamics reported in the literature: POD mode 2 is related to the very low frequency asymmetric or deviation global mode, modes 3 and 4 to vortex shedding with oscillation in the horizontal direction, modes 5 and 6 to vortex shedding with oscillations in the vertical 
direction, and modes 7 and 8 to (symmetric) wake pumping.

\begin{figure}[h]
\hspace{-3in}
\begin{tabular}{cc}
\includegraphics[height=65mm]{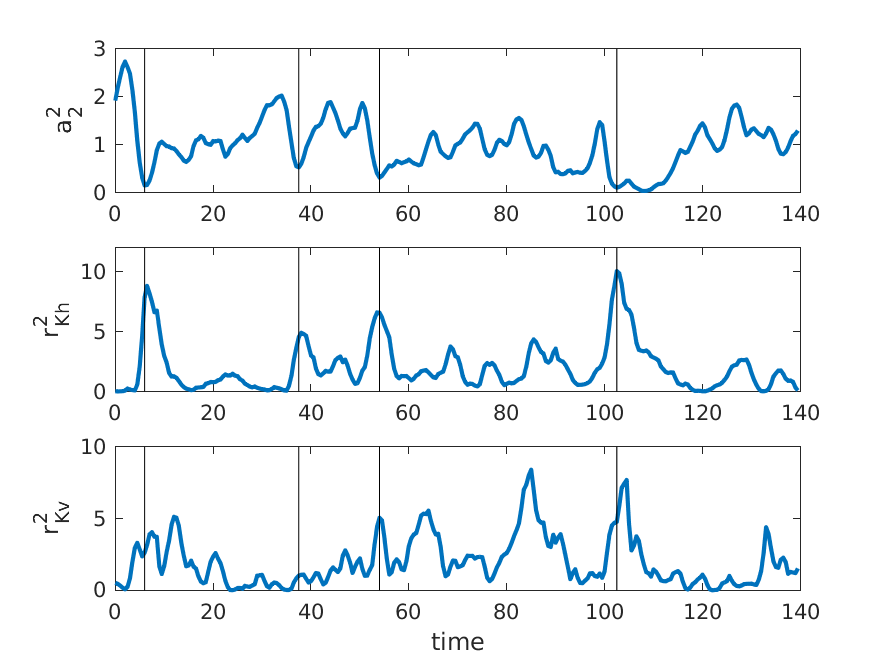} &
\includegraphics[height=65mm]{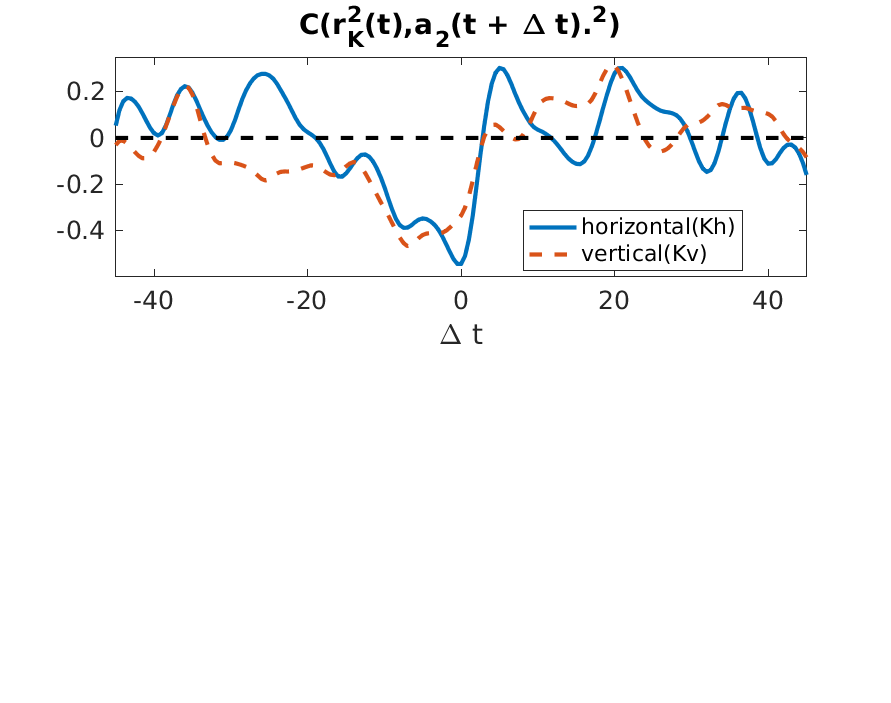} \\
\end{tabular}
\caption{ 
Left) energy of POD modes corresponding to
quasi-steady deviation $a_2^2$, spanwise 
($r_{Kh}^2=a_3^2+a_4^2$) and vertical ($r_{Kv}^2=a_5^2+a_6^2$)  
vortex shedding intensities -
vertical lines correspond to minima of $a_2^2$;
right) correlation coefficient between  vortex shedding energy and quasi-steady deviation energy. 
}
\label{fig16:correla2}
\end{figure}

The POD decomposition allows to investigate quantitatively the correlations between the global modes during the wake dynamics. We first 
compare the intensity of the deviation global mode given by ($a_2^2$)
with the magnitude of vortex shedding in the horizontal (spanwise)
and  the vertical direction which can respectively be obtained 
with the sums of the amplitudes $r_{Kh}^2=a_3^2+a_4^2$ and $r_{Kv}^2=a_5^2+a_6^2$.
We can see in figure ~\ref{fig16:correla2} that minima of $a_2$ are associated with high
energy in spanwise vortex shedding. The correlation coefficient between
$|a_2|$ and ${|a_3|^2+|a_4|^2}$ is strongly 
negative (-0.6).
Due to the three-dimensional nature of vortex shedding, 
spanwise and vertical shedding motions are correlated (0.48),
so that a negative correlation is also obtained  between $|a_2|$ and $a_5^2+a_6^2$ (-0.33). 
The negative correlation is consistent with the idea that a reduced
asymmetry is accompanied by an increase in vortex shedding,
which was observed in control experiments of \cite{kn:brackston16} and \cite{kn:li16}.

\begin{figure}[h]
\hspace{-3in}
\begin{tabular}{cc}
\includegraphics[height=65mm]{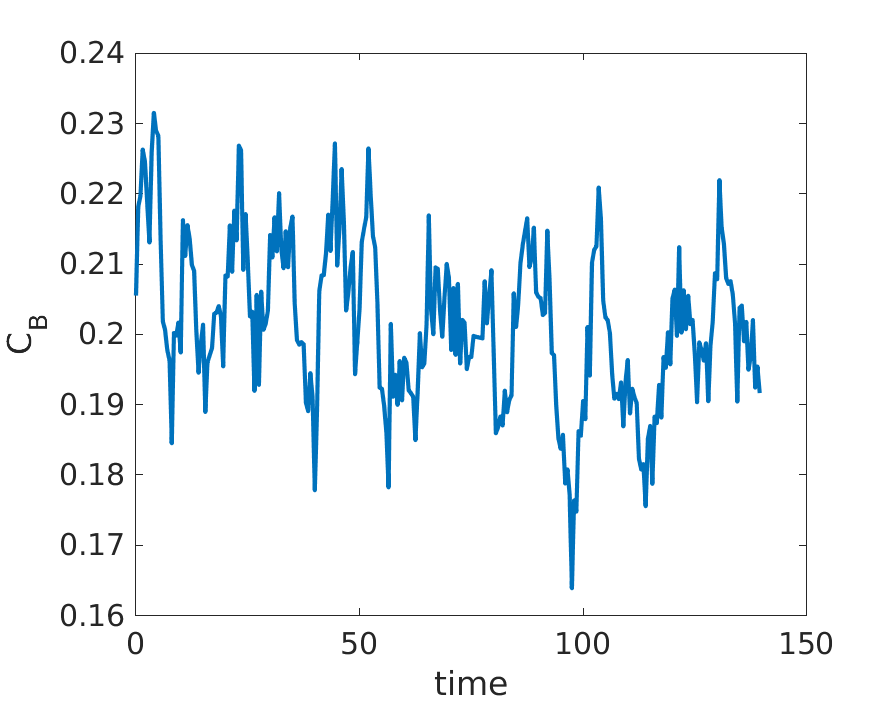} 
\begin{minipage}{0.5\textwidth}
\includegraphics[height=65mm]{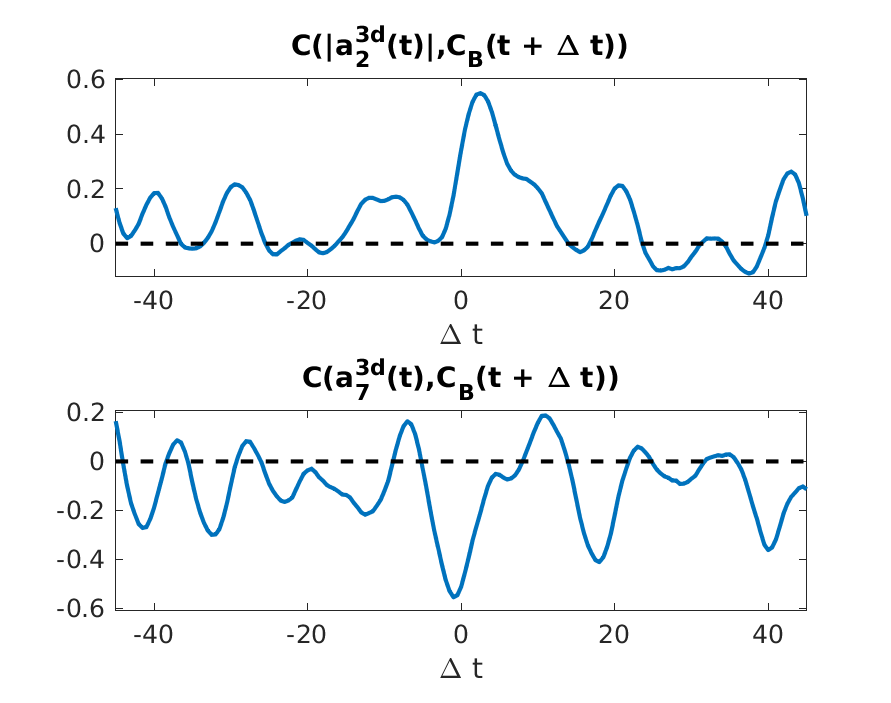} 
\hspace{-1cm} \includegraphics[height=65mm]{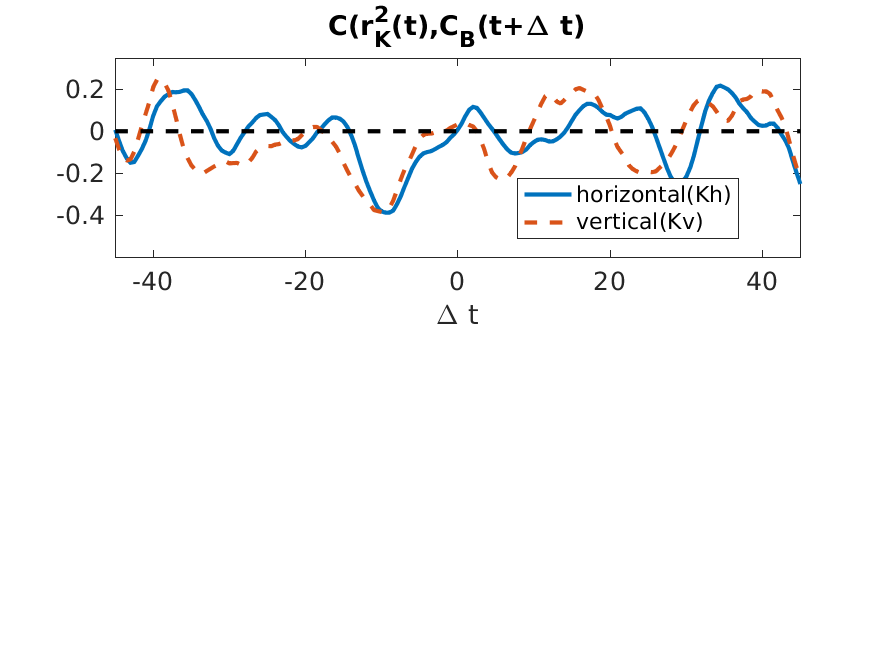} 
\end{minipage} \\
\end{tabular}
\caption{
Left) base suction  coefficient $C_B=-C_p$ obtained by integrating the pressure
over the rear of the body;
right) 
 correlation between the base suction coefficient  and the 
POD amplitudes corresponding to the steady deviation ($a_2$, top), wake pumping mode  
($a_7$, middle) and vortex shedding modes ($r_{Kh,v}^2$, bottom).} 

\label{fig17:correlcb}
\end{figure}

 The next step is to compare the evolution of the POD amplitudes 
with the  base suction coefficient $C_B=-C_{pb}$ where the 
pressure coefficient $C_{pb}$ corresponds to the integral of 
the pressure over the base of 
the body, which is shown in figure ~\ref{fig17:correlcb} (left).  
Figure ~\ref{fig17:correlcb} (right)  shows that
the base suction coefficient is positively
correlated with the amplitude of the wake deviation $a_2^{3d}$
with a positive delay of $\Delta t U/H \sim 2.5$ and 
with a maximum correlation coefficient of 0.55 (
we note that the 
correlation coefficient with the 2-D deviation mode $a_2^{2d}$ is 0.41. 
with a time delay of 4 time units).
The time delay appears to have some significance as the correlation without time delay drops to 0.34.
This suggests that the variations of the drag follow those of the mean deviation amplitude.
This is in agreement with the idea that a drag reduction corresponds to 
a decrease in the quasi-steady  wake asymmetry, as was observed by \cite{kn:cadot16} and \cite{kn:pavia18}.
Figure ~\ref{fig17:correlcb} also shows that
a correlation coefficient of -0.55 with a negative delay of $\Delta t U/H \sim -1.5$ is observed between
the pressure and the amplitude of $a_7$.
Unlike the previous case, the correlation remains about the same without no time delay (-0.51), so it
is not possible to assign a physical relevance to the small time delay observed.
As seen above, $a_7$  is associated with wake 
pumping: from figure ~\ref{fig14:modes310} one can see
in the near-wake that when $a_7 >0$
there are more negative fluctuations within the zone, so that
the size of the recirculation actually increases.
This is consistent with the observation that the base suction 
coefficient decreases as the recirculation length increases.

In contrast, the correlation of the drag with the intensity of vortex shedding appears
to be weaker (and negative).

The observations made above suggest the following picture: 
the flow is characterized by a quasi-steady wake deviation, vortex shedding modes and
low-frequency wake pumping modes.
The drag coefficient depends on the global (symmetric) size of the bubble,
which is associated with  wake pumping, as a longer bubble corresponds to a lower drag.
It  also  depends on the magnitude  of the quasi-steady wake asymmetry:
a lower drag corresponds to a decrease of asymmetry, and to an increase
in the intensity of vortex shedding,  
 a negative correlation of vortex shedding in  spanwise and vertical directions
with both the quasi-steady deviation and the base suction coefficient.

\section{ Low-dimensional model}

We now examine whether it is possible to model the behavior of the largest scales using
a POD-based model, even if the scales considered represent only a fraction of 
the total fluctuating kinetic energy. 
Following the approach described in \cite{kn:pf09}, \cite{kn:pre17}, we build a low-dimensional model
to reproduce the dynamics. We use a Galerkin approach to project the Navier-Stokes equations
onto the basis of spatial modes for a selected truncation, and obtain a set of ordinary differential equations for the 
normalized amplitudes $a_n(t)$.
The model is of the form

\begin{equation}
\dot{a}_n= L_{nm} a_m + Q_{nmp} a_m a_p + T_{n}.
\label{evoleqn}
\end{equation}
where
\begin{itemize}
\item  the linear terms contain the viscous dissipation
\begin{equation}
L_{nm}= \int \nu \Delta \pphi_{m} . \pphi_{n} d\xx.  
\end{equation}
For the ten-mode truncation they form  a diagonal matrix $L \sim -0.05 I$
as indicated in table ~\ref{tab:coef}.
\item  $T_n$ is a closure term representing the contribution of the unresolved stresses (associated 
with the modes excluded from the truncation) to the evolution of the 
amplitude $a_n$.

\item  the quadratic terms $Q_{nmp}$ can be written in symmetric form as
\begin{equation}
Q_{nmp}= \frac{\sqrt{\lambda_m \lambda_p}}{\lambda_n} 
\frac{1}{2}(2-\delta_{mp}) \int (\pphi_p . \nabla \pphi_{m}) + \pphi_m . \nabla \pphi_p) . \pphi_{n} d\xx.  
\end{equation}

For the evolution equation of  the amplitude $a_n$, $1 < n \le 10$,
shown in equation (~\ref{evoleqn}),
the interaction coefficient of $a_n$ with the mean mode, $Q_{nn1}$, 
is essentially independent of $n$ and its value is about 0.2.   
A physical interpretation of this is that each of the modes interacts 
directly and equally with the mean mode, in particular 
its strong shear layers.

\end{itemize}

Generally speaking, the magnitudes of the quadratic coefficients $Q_{nmp}$ 
provide insight into the interactions between the different modes.
Table ~\ref{tab:quadratic} contains the values of the quadratic coefficients $Q_{nm1}$
 for $3 \le n,m \le 10$.
The dominant values which will be kept for the model are indicated in bold. 

\subsection{Simplified model }

We first consider a simplified version of the model 
by making the following assumptions:
\begin{itemize}
\item we assume that the first two modes $a_1$ and $a_2$ are constant.

\item we neglect quadratic terms of small magnitude.

\item we model the energy transfer to the unresolved modes
by assuming that their effect is to compensate for the production term
i.e the interaction with the mean shear (mode 1). This means that
\begin{equation}
T_n =  (- L_n - Q_{1nn} a_1 ) a_n.
\end{equation}

\end{itemize}
This leads to the following form  for the model, which we will refer to as $M_0$: 
\begin{eqnarray} 
\dot{a}_1 &= & 0 \\
\dot{a}_2 &= & 0 \\
\dot{a}_3 &=& 0.91 a_4  \\
\dot{a}_4 &=& -0.94 a_3  \\
\dot{a}_5 &=& -1.09 a_6  +  0.49 a_9 \\
\dot{a}_6 &=& 1.22 a_5  \\
\dot{a}_7 &=& -0.42 a_8   \\
\dot{a}_8 &=& -0.55 a_5 +0.44 a_7  \\
\dot{a}_9 &=& -0.6 a_4  -0.86 a_{10} \\
\dot{a}_{10} &=& -0.88 a_3 + 0.98 a_9   
\end{eqnarray}

We can see that there are only interactions between modes
with the same parity: (3,4,9,10) on the one hand and (5,6,7,8)
on the other hand.
The interactions between the normalized 
amplitudes $(a_{2i-1},a_{2i}), 2 \le i \le 5$
are of the form
$ a_{2i-1}=-\omega_i (\frac{\lambda_{2i}}{\lambda_{2i-1}})^{1/2} a_{2i}$, 
$ a_{2i}=\omega_i (\frac{\lambda_{2i-1}}{\lambda_{2i}})^{1/2} a_{2i-1}$, 
which correspond to \berengere{propagative} oscillatory solutions for the amplitudes
$\sqrt{\lambda_n} a_n$, in agreement
with the convective dynamics expected for the corresponding 
\berengere{K\'arm\'an} modes.

The averaged frequencies $|\omega_i| = 2 \pi f_i$ 
identified for the pairs (3,4), (5,6), (7,8) and (9,10)
are  0.92, 1.15, 0.43, 0.9, which correspond to frequencies (or Strouhal numbers) of
0.15, 0.18, 0.07 and 0.14.  
This is in  good agreement with the main 
frequencies identified in the simulation.
We emphasize that this prediction of the relevant time scales is based  
exclusively on the spatial structure of the modes, 
which are extracted from a set of samples arbitrarily separated in time.
Since the computation is based on the derivatives of the spatial modes,
some uncertainty exists in the determination of the time scales.

\begin{figure}[h]
\hspace{-3in}
\includegraphics[height=75mm]{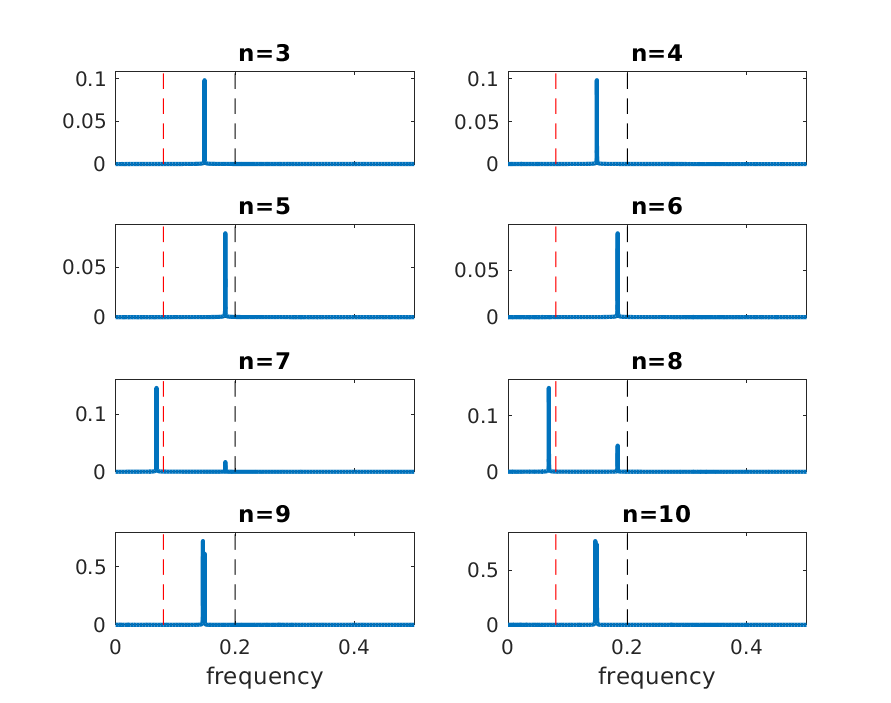} 
\caption{Power spectral density of the 3-D POD mode amplitudes 3 to 10
$|\hat{a}_n^{3D, M_0}|^2$ in the simplified model 
- the red and black lines respectively correspond to the two 
frequencies 0.08  and 0.2.} 
\label{fig18:af3dmodel}
\end{figure}

The model was integrated from a random initial condition,
and the amplitudes $a_i, 3 \le i \le 10$ are 
represented in spectral space in figure ~\ref{fig18:af3dmodel}.
As expected, the frequencies of the model coefficients
agree well with  the dominant frequencies
identified in the previous section for the 
amplitudes of the modes in the simulation. 
The amplitudes of the modes are also close to their expected values, as shown in table ~\ref{tab:coef}.
These results indicate that the main temporal dynamics of the flow 
can be recovered from the quadratic interactions between spatial POD modes,
even if the snapshots are obtained with large separations, which is
evidence of the predictive abilities of the POD-based model.

\subsection{Switches}

We now examine if and how switches can be reproduced by a more
complex version of the model, which we derive by  
relaxing the  assumptions of the simplified model as follows:
\begin{itemize}
\item the second mode is allowed to vary.  
\item feedback is provided between the unresolved terms and the modes of the truncation. 
\end{itemize}
As shown in ~\cite{kn:pre17}, 
we assume that the rate of energy transferred to the small scales
depends on the energy available in the largest scales.
If more  energy is available in the large scales, then more energy is extracted
by the small scales and conversely.
This leads to us to introduce a time-varying viscosity term, so
that the effect of the unresolved terms is modelled as 
\begin{equation}
T_n  = A_n a_n + \epsilon_n 
\label{tn}
\end{equation}
where
\begin{enumerate}
\item  the time-averaged value $<A_n>$ satisfies 
\begin{equation}
 <A_n>   = -L_n - Q_{nn1},
\end{equation}
\item  $A_n$ contains a linear part and a quadratic part 
\begin{equation}
 A_n   = <A_n>  -  \alpha_n \sum_{p= 2}^{10} \lambda_p  +  \alpha_n \sum_{p= 2}^{10} \lambda_p |a_p|^2 ,
\end{equation}

As has been shown in \cite{kn:pre17}, we have
\begin{equation}
T_n =  (<A_n> + \alpha_n \sum_{p \ge 1}^{N} \lambda_p - \alpha_n \sum_{p \ge 1}^{N} |a_p|^2 ) a_n + \epsilon_n
\end{equation}
where 
\[ \alpha_n= -\frac{<A_n>}{2 \sum_{p \ge 2}^{N} \lambda_p}. \]

 The value of $\alpha_i$  was evaluated  
to be around 0.5. 
Since the model was not found to be largely insensitive to 
the exact values of $\alpha_i$, in what follows 
a constant value of $\alpha_i = \alpha =0.5$ was used 
for all modes for the sake of simplicity. 
Examination of equation (\ref{tn}) 
and of the POD eigenvalues shows that about 50\%
of the turbulent viscosity is dependent on the amplitude of mode $a_2$.
In that sense the structure of the model displays similarities with 
both \cite{kn:rigas15_a} and \cite{kn:cadot15}'s model,
which includes a cubic term in $a_2$.
However it is derived from  entirely different physical arguments. 

\item $\epsilon_n$ represents Gaussian noise.
The amplitude of the noise  used to integrate the model 
was determined using  $\epsilon_n \approx |A_n|$.
We used $\epsilon_2=0.09$
and $\epsilon_i=0.04$ for $3 \le i \le 10$. 
[A Gaussian variable was generated every $2.5$ convective time units
and linearly interpolated was used for the integration].

\end{enumerate}

The modified model reads: 
\begin{eqnarray} 
\dot{a}_1 &= & 0 \\
\dot{a}_2 &=& (L_2 - \alpha \sum_{2}^{10} a_p^2 \lambda_p)  a_2 + \epsilon_2  \\
\dot{a}_3 &=& (L_3 - \alpha \sum_{2}^{10} a_p^2 \lambda_p)  a_2  + 0.71 a_4 + \epsilon_3   \\
\dot{a}_4 &=& (L_4 - \alpha \sum_{2}^{10} a_p^2 \lambda_p)  a_2  -0.74 a_3  + \epsilon_4 \\
\dot{a}_5 &=& (L_5 - \alpha \sum_{2}^{10} a_p^2 \lambda_p)  a_2  -1.09 a_6   + \epsilon_5 \\
\dot{a}_6 &=& (L_6 - \alpha \sum_{2}^{10} a_p^2 \lambda_p)  a_2  -1.2 a_5  + \epsilon_6 \\
\dot{a}_7 &=& (L_7 - \alpha \sum_{2}^{10} a_p^2 \lambda_p)  a_2  -0.42 a_8  + \epsilon_7  \\
\dot{a}_8 &=& (L_8 - \alpha \sum_{2}^{10} a_p^2 \lambda_p)  a_2  -0.55 a_5 +0.44 a_7  + \epsilon_8 \\
\dot{a}_9 &=& (L_9 - \alpha \sum_{2}^{10} a_p^2 \lambda_p)  a_2  -0.6 a_4  -0.86 a_{10} + \epsilon_9 \\
\dot{a}_{10} &=& (L_{10} - \alpha \sum_{2}^{10} a_p^2 \lambda_p)  a_2  -0.88  a_3  + 0.98 a_9 + \epsilon_{10}.  
\end{eqnarray} 

The quadratic terms in the expression for $a_2$ 
are not included in the model as their magnitude was small
(we checked that including  these terms in the equations did not change the dynamics reported below). 

The effect of the feedback term on the dynamics of the model is that if there is less energy in mode $a_2$,
the higher-order modes will extract less energy, which will allow mode $a_2$ to grow.
Conversely, if mode $a_2$ becomes too large, the energy transfer to the higher-order modes will be increased,
which will in turn affect the energy of mode $a_2$. 

\begin{figure}[h]
\hspace{-2in}
\begin{minipage}{0.5 \textwidth}
\includegraphics[height=55mm]{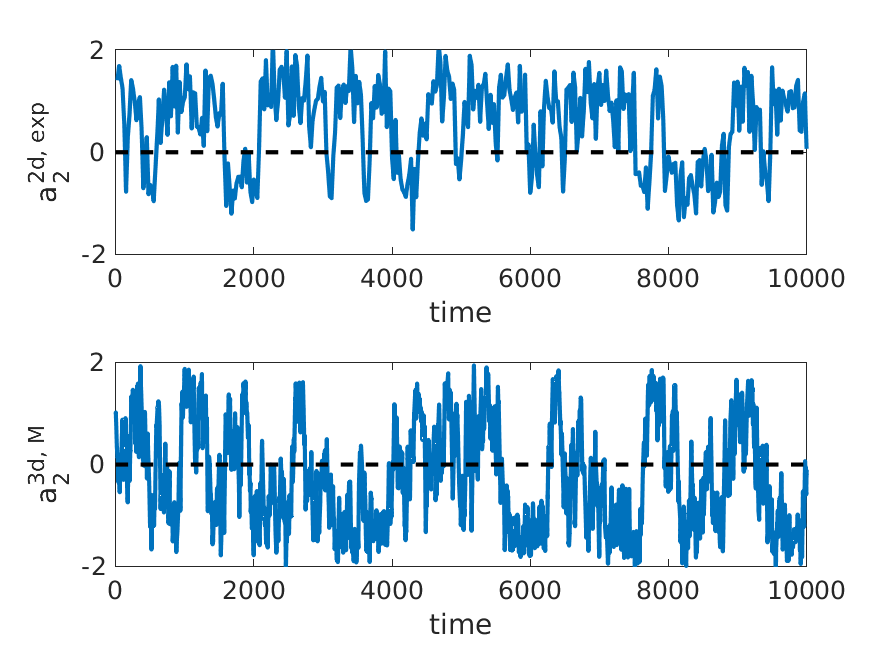}
\end{minipage}
\begin{minipage}{0.5 \textwidth}
\includegraphics[height=50mm]{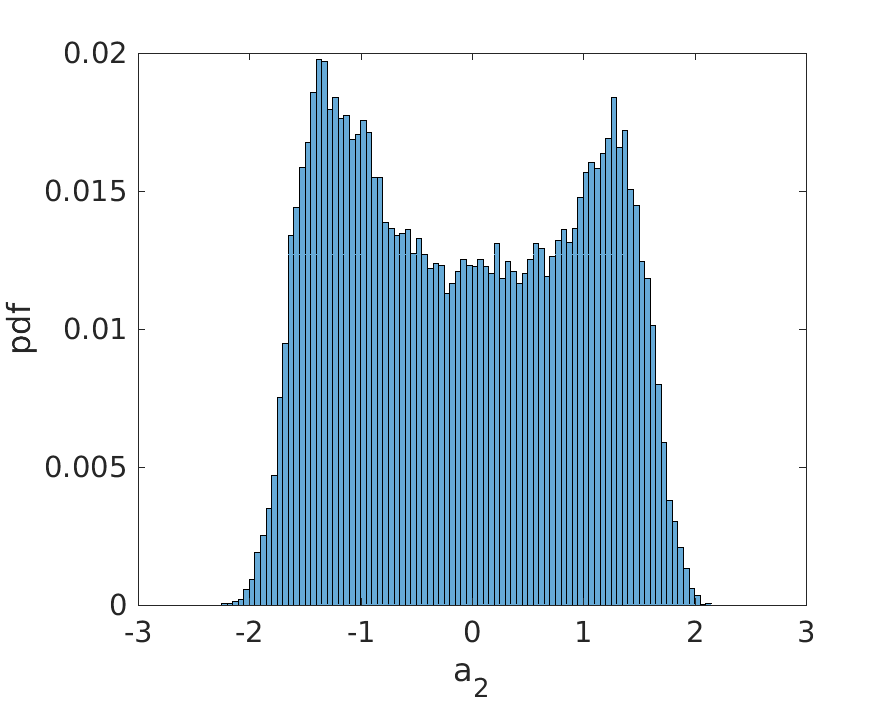}
\end{minipage}
\caption{ Left) POD amplitude $a_2$; top: model M (3D); 
bottom: experiment (2D);
right) histogram of $a_2$
} 
\label{fig19:a2compmodelexp}
\end{figure}

\begin{figure}[h]
\hspace{-2in}
\begin{minipage}{0.5 \textwidth}
\hspace{-1cm} \includegraphics[height=62mm]{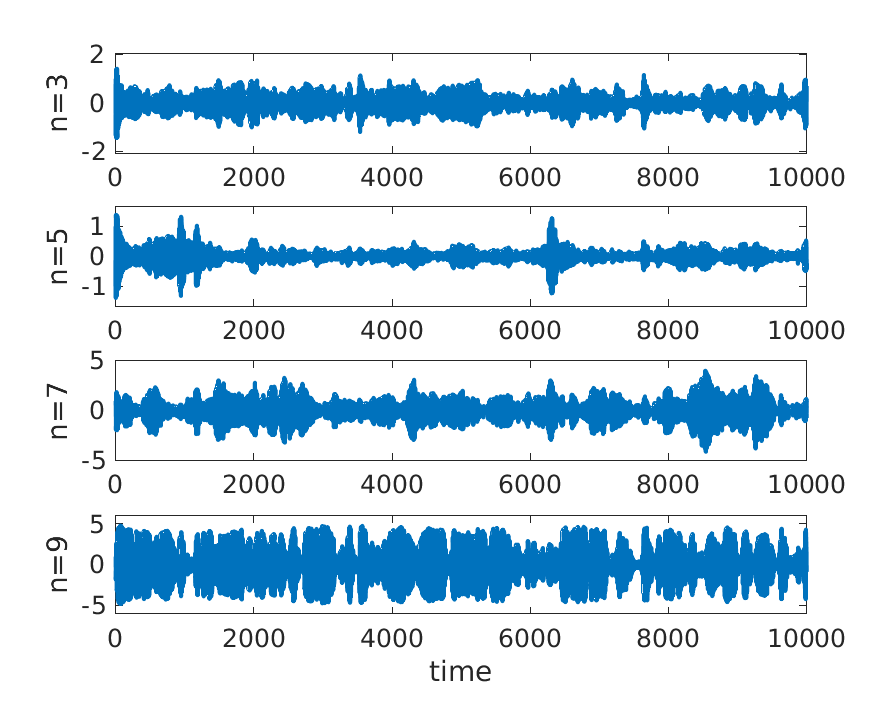}
\end{minipage}
\begin{minipage}{0.5 \textwidth}
\includegraphics[height=65mm]{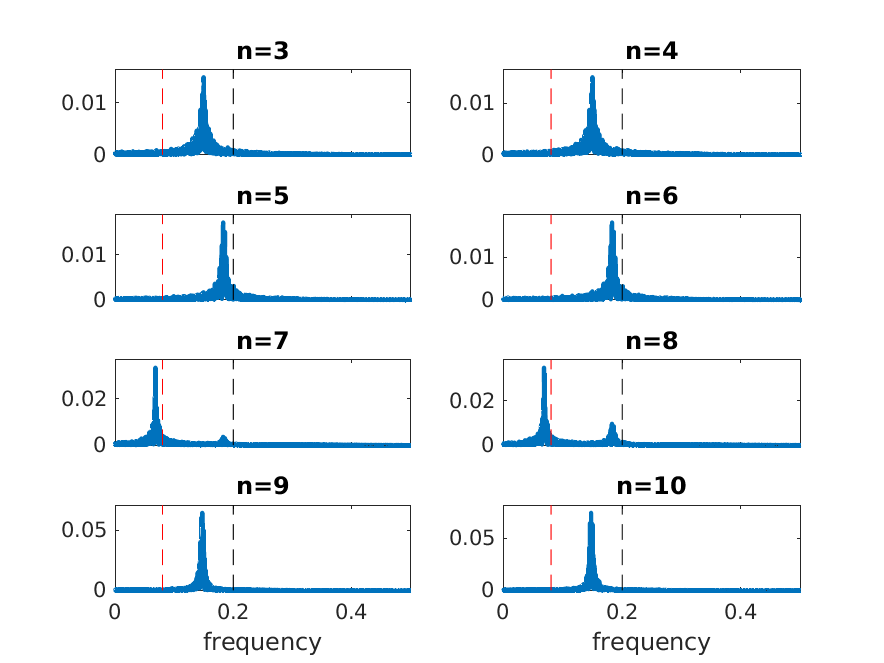}
\end{minipage}
\caption{ 
Left) amplitudes of 3-D POD amplitudes of modes 3, 5 , 7 and 9  in the model;
right) power spectral density of POD amplitudes in the model $|\hat{a}_n^{3d,M}|^2$
- the red and black lines respectively correspond to the two 
frequencies 0.08  and 0.2.} 
\label{fig20:afmodel}
\end{figure}

Figures ~\ref{fig19:a2compmodelexp} and ~\ref{fig20:afmodel} 
shows results of the model integration for a noise amplitude of about 0.15.
Figure ~\ref{fig19:a2compmodelexp}  (left) shows
the 3-D coefficient $a_2^{3d}$ predicted with the model M along with
the 2-D coefficient $a_2^{2d}$ in the experiment,
which appears a relevant comparison since, as shown   
in the previous section, there is a reasonably good correlation between
$a_2$ in 3-D and 2-D (0.6).
The  model displays time scales of $O(1000)$ 
between switches, in agreement with experimental observations. 
Figure ~\ref{fig19:a2compmodelexp} (right) shows that the histogram of the amplitude is similar
to that observed in the experiment in figure ~\ref{fig6:hista14}.
This shows that the model is able to reproduce deviations in a way that is consistent with experiments.
As figure ~\ref{fig20:afmodel} indicates, modes 3 to 10 are characterized by relatively 
fast oscillating time scales and slower amplitude variations.  
The frequencies of the amplitudes are shown in figure ~\ref{fig20:afmodel} (right) and compare relatively 
well with those measured in the simulation, given the crudeness of the truncation.
Table ~\ref{tab:coef} (last line) shows that the  magnitude
of the normalized POD amplitudes  is relatively well estimated by the model
with  values of about 0.2 to 2 for the last modes
of the truncation (we emphasize that the model contains only 20\% of the total fluctuating energy).
The main dynamics of the largest scales are therefore captured by the model. 

\begin{table}
\begin{tabular}{ccccccccccc}
n & 1 & 2 &3 & 4 & 5 & 6 & 7 & 8 & 9 & 10 \\
$L_i$ & & -0.05 & -0.05 &-0.05  &-0.05  &-0.05  &-0.05  &-0.05  &-0.05  & -0.05 \\
$<a_n^2>_{M_0}$ & 1 & 1 & 0.93 & 0.93 & 0.86 & 0.86 & 1.45 & 1.51  & 1.05 & 1.05 \\
$<a_n^2>_{M}$ & 1 & 1.05 & 0.29 & 0.29 & 0.18 & 0.18 & 1.01 &1.03  & 2.05 & 2.33 \\
\end{tabular}
\caption{ Model linear coefficients and predicted energy}
\label{tab:coef}
\end{table}

\begin{table}[h]
\begin{tabular}{ccccccccc}
$Q_{1jm}$ & m=3 & m=4 & m=5 & m=6 & m=7 & m=8 & m=9 & m=10 \\
j=3 & 0.21  & {\bf 0.91}  & & & & -0.22 & & 0.37  \\ 
j=4 & {\bf -0.94}  & 0.23  & & & & 0.0 & 0.29 & 0.17  \\ 
j=5 &   &   & 0.22 & {\bf 1.09} & 0. &  0.34 &  &   \\ 
j=6 &   &   & {\bf -1.22} & 0.20 & -0.12 & -0.10  &  &   \\ 
j=7 &   &  &  & 0.17 & 0.16 & {\bf -0.42} &  &   \\ 
j=8 &   &   &  {\bf 0.55}  & & {\bf 0.4}  & 0.17  &  &   \\ 
j=9 & 0.32   &   & {\bf 0.59} &  &  &  & 0.22 & {\bf -0.86}   \\ 
j=10 & {\bf -0.88}   &  -0.18  &  &  &  &  &{\bf 0.98}   & 0.24    \\ 
\end{tabular}
\caption{Quadratic interaction coefficients with the main mode $Q_{j1m}$; Only coefficients
larger than 0.1 are indicated; only coefficients larger than 0.4 (indicated in bold) are 
included in the model. }
\label{tab:quadratic}
\end{table}

\section{Conclusion}

We have applied Proper Orthogonal Decomposition to the 3-D numerical simulation
of the flow behind an Ahmed body at $Re= 10^4 $.
Reflection symmetry was applied to the computed data set  in order to compensate
for the relatively short time of the simulation, which precludes the observation of switches
in the wake deviation. 
As a consequence of the enforced statistical symmetry, the flow can be 
decomposed into symmetric and antisymmetric structures.
The mean flow consists of a symmetric recirculation bubble and an 
antisymmetric deviation, 
the effect of which is to gather flow streamlines 
around one of the base diagonals.
2-D POD analysis  was performed in the near-wake in the simulation and compared
with experimental results obtained for the same geometry. 
Despite the discrepancy in Reynolds number between the simulation and the experiment,  
an excellent agreement was observed for both the spatial structure and temporal
statistics of the POD modes. 

2-D results were then confronted with a 3-D approach.
The energetic importance of the quasi-steady wake deviation was established.
The evolution of this global 3-D  deviation mode 
was relatively well captured by  2-D measurements  in the near-wake.
The next most energetic patterns are associated with vortex shedding and
wake pumping.
Characteristic frequencies were identified for each type of structure.
Structures associated with  wake pumping were characterized by a low
frequency of about 0.08. 
Both symmetric and antisymmetric structures associated with 
vortex shedding in respectively the vertical and spanwise direction were 
characterized by   dominant frequencies of  0.19 and 0.23 in the far wake.
A strong negative correlation  was noted between the intensity of the 
vortex shedding modes and the magnitude of the quasi-steady deviation. 
 In addition,
increases in the base drag were found to correspond to an increase  
of the quasi-steady deviation, along with a shrinkage of the recirculation
zone associated with wake pumping.

Finally, POD-based low-dimensional models were derived for the largest scales of the flow.
The energy content of the modes was correctly captured by the model.
A simplified model was able to single out the  main  frequencies of the POD  amplitudes 
observed in the simulation from the spatial modes, regardless of the time 
separation between the snapshots used to compute POD. 
This predictive ability of the model is remarkable in view of  
the slow convergence of the decomposition, which reflects the complexity of the flow.
A more elaborate version of the  model was also considered.
The approach is consistent with Rigas et al.'s model \cite{kn:rigas15_a} and in particular 
the structure of the model would be the same if the POD truncation was limited to two modes. 
By accounting for the effect of the unresolved modes
with a feedback term, the POD-based 
model was able to reproduce the characteristics of the
switches in the wake deviation. 
The success of the model supports the idea that wake switching is triggered by higher-order modes.

\bibliographystyle{plain}
\bibliography{all}

\pagebreak

\pagebreak

\end{document}